\documentclass[letterpaper,onecolumn,prx,aps,showpacs,superscriptaddress,longbibliography,floatfix]{revtex4-2}

\usepackage{amssymb, amsmath, bbm, amsthm, changes, graphicx, dsfont, verbatim, appendix, physics} 

\usepackage[hyperindex, breaklinks]{hyperref}
\hypersetup{
     colorlinks=true,        
     citecolor=blue,    
     filecolor=blue,      		
     urlcolor=blue,           	
    runcolor=cyan,
}

\usepackage{enumitem}   
\usepackage{xcolor}
\usepackage{orcidlink}
\usetikzlibrary{decorations.pathreplacing}

\def\Z2{\mathbb Z_2}

\def\b|{\big \|}
\def\dist{\textrm{dist}}
\def\supp{\textrm{supp}}
\def\om{\omega}
\def\Eom{\underset{\omega}{\mathbb{E}} }

\theoremstyle{definition}
\newtheorem{Lemma}{Lemma}

\newtheorem{Theorem}[Lemma]{Theorem}

\begin{document}
\title{Stability of slow Hamiltonian dynamics from Lieb-Robinson bounds}

\author{Daniele Toniolo\,\orcidlink{0000-0003-2517-0770}}
\email[]{d.toniolo@ucl.ac.uk}
\email[]{danielet@alumni.ntnu.no}
\affiliation{Department of Physics and Astronomy, University College London, Gower Street, London WC1E 6BT, United
Kingdom}

\author{Sougato Bose\,\orcidlink{0000-0001-8726-0566}}
\affiliation{Department of Physics and Astronomy, University College London, Gower Street, London WC1E 6BT, United
Kingdom}


\begin{abstract}
We rigorously show that a local spin system giving rise to a slow Hamiltonian dynamics is 
stable against generic, even time-dependent, local perturbations. The sum of these perturbations can cover a significant amount of the system's size. The stability of the slow dynamics follows from proving that the Lieb-Robinson bound for the dynamics of the total Hamiltonian is the sum of two contributions: the Lieb-Robinson bound of the unperturbed dynamics and an additional term coming from the Lieb-Robinson bound of the perturbations with respect to the unperturbed Hamiltonian. Our results are particularly relevant in the context of the study of the stability of Many-Body-Localized systems, implying that if a so called ergodic region is present in the system, to spread across a certain distance it takes a time proportional to the exponential of such distance. The non-perturbative nature of our result allows us to develop a dual description of the dynamics of a system. As a consequence we are able to prove that the presence of a region of disorder in a ergodic system implies the slowing down of the dynamics in the vicinity of that region.
\end{abstract}

\maketitle



The Lieb-Robinson (L-R) bounds quantify, in non-relativistic quantum physics, the maximal speed at which an operator, that initially acts locally, spreads its action over the system as a function of time because of the Hamiltonian dynamics. This introduces in this context the concept of the lightcone: the effect of the spread of the support of the operator can be detected at a certain distance only after a certain time interval, before that it is exponentially small. The pioneering work of Lieb and Robinson \cite{Lieb_Robinson}, has been improved, in particular by Hastings at the beginning of the 2000's \cite{Hastings_2004}, with applications nowadays ranging from condensed matter theory \cite{Hastings_2014} to quantum information and simulation  \cite{Bravyi_Hastings_Verstraete_2006,Osborne_2007} and quantum chaos. The recent reviews \cite{Gong_2022, Lucas_review} collect some of these works. We also refer to \cite{Nach_2019} for the mathematical foundations of the L-R bounds.

In this work we evaluate the L-R bounds of a one-dimensional system with total Hamiltonian $ H + \sum_j h_j $ starting from the L-R bounds of the local Hamiltonian $ H $ when $ \sum_j h_j $ is an additional local Hamiltonian, that we call perturbation, but that can cover a portion of the system comparable with its size and can have intensity as large as that of $ H $.

If a one dimensional lattice Hamiltonian gives rise to a slow dynamics, as quantified by equation \eqref{LR_bound_MBL}, our theorem \ref{0} implies that the dynamics $ H + \sum_j h_j $ continues to be slow. The meaning of our result is very transparent considering the case in which the system hosts a single perturbation \eqref{bound_single_slow}. The spread of the perturbation $ h $, given by the Heisenberg dynamics of the full Hamiltonian $ H+h$, happens with a time scale that is exponentially long with the distance, as in the unperturbed case with the Hamiltonian $ H $.

Our result, theorem \ref{0}, is {\it non-perturbative} and takes into account all possible kinds of dynamics arising from $ H $. In fact theorem \ref{0} allows us to develop a dual description of the dynamics of a system: when there are a pair of Hamiltonians $ H $ and $ H' $, and perturbations $ \sum_j h_j $, $ \sum_j h_j' $, such that $ H + \sum_j h_j = H' + \sum_j h_j' $, then if $ H $ or $ H' $ gives rise to a slow dynamics then the dynamics of the full Hamiltonian is, locally, slow. This implies, for example, that a region of disorder in a ergodic system  slows down the dynamics in the vicinity of that region.

The motivation of our work arises, in the context of the study of Many-Body-Localized (MBL) phases of matter \cite{Basko_2006, Gornyi_2005, Oganesyan_2007, Znidaric_2008}, see also the reviews \cite{Abanin_2019, Sierant_2024}, from the the long term debate spurred by the papers \cite{De_Roeck_2017_1,Luitz_2017}, regarding the stability of MBL with respect to rare ergodic regions. Our result \eqref{bound_single_slow} states that such ergodic regions, within a phase of slow dynamics, propagate at most exponentially slowly, namely to spread across a certain distance they take a time proportional to the exponential of such a distance. We also discuss how regions of anomalous slow dynamics inside an ergodic system slow down the dynamics in the vicinity of such regions.

\section{Definition of the slow Hamiltonian dynamics by Lieb-Robinson bounds}

With slow dynamics we precisely mean the class of systems with a L-R bound giving rise to a logarithmic lightcone, defined as follows. Let us consider the operators $ A $ and $ B $ with supports on the regions $ \supp(A) $ and $ \supp(B) $ of the lattice, this means that outside $ \supp(A) $, respectively $ \supp(B) $, they act like the identity. We assume for simplicity that these supports are simply connected. $ H_\om $ gives rise to a slow dynamics when it holds, with $\beta>0$:
 \begin{align} \label{LR_bound_MBL}
  \underset{\omega}{\mathbb{E}}  \b| [e^{iH_\omega t} A e^{-iH_\omega t},B] \b| \le   \, K \, \|A\| \, \|B\|   \,  t^\beta \, e^{-\frac{\dist(\supp(A), \supp(B)}{\xi} }
 \end{align}
With the symbol $ \underset{\omega}{\mathbb{E}}(\cdot) $ we denote averaging with respect to all the possible realizations of the Hamiltonian $ H_\omega $, that, for disordered systems, correspond to the different realizations of disorder $  \omega = \{\omega_j\}_{j\in \Lambda} $. $\Lambda$ denotes the lattice. 
This type of L-R bounds have been proven for one-dimensional many-body localized systems in the context of the so called Local-Integral-of-Motion model \cite{Kim_2014,Nach_2021,Zeng_2023,Toniolo_2024_2}, with $ \beta=1 $, and recently, despite only for low energies, from a rigorous analysis of the $ XXZ $ model with random magnetic in \cite{Elgart_2023}. The bound \eqref{LR_bound_MBL} in the remarkable work  \cite{Elgart_2023} unfortunately suffers from a $ K $ that is system's size dependent. In all the following $ K $ is assumed to be system's size independent. 
 The bound \eqref{LR_bound_MBL} could in principle be given from a deterministic Hamiltonian, potential candidates are disorder-free systems arising in the context of gauge theory like \cite{Smith_2017,Smith_2019}, and Hamiltonians in relation the so called ”surface codes“, according to \cite{Yin_2024} that showed the existence of exponentially localized low energy eigenstates in a model of that kind. Nevertheless the rigorous derivation of \eqref{LR_bound_MBL} in a deterministic setting is unknown to us.

\section{Our results on the stability of the slow Hamiltonian dynamics}

The name ``logarithmic lightcone'' comes from the equality $ K \, t^\beta \, e^{  -\frac{d}{\xi} } =   e^{\log(Kt^\beta)  -\frac{d}{\xi} } $, where we have defined $ d:=\dist(\supp(A), \supp(B)$. $ K $ is  such that $ K \, t^\beta $ is dimensionless. This means that up to a time $ t $ exponentially large in $ d $,  $ e^{iH_\om t} A e^{-iH_\om t} $ and $ B $ commute up to an exponentially small correction. In fact defining $ K t_{max}^\beta = e^{\frac{d}{2\xi}} $, at $ t=t_{max} $ the upper bound in \eqref{LR_bound_MBL} is proportional to $ e^{-\frac{d}{2\xi}} \ll 1 $, when $ d \gg \xi $. $ e^{iH_\om t} A e^{-iH_\om t} $ is a propagating operator and if we imagine $ B $ as a ``detector'' of such propagation, according to \eqref{LR_bound_MBL} within the time $ t_{max} $ there is virtually no detection.

We now informally state our theorem \ref{0}, according to equation \eqref{bound_single}, in the case of a single time-independent perturbation $ h_\om $ acting on a system with the L-R bounds \eqref{LR_bound_MBL}. $ h_\om $ can be a random matrix, modelling a so called ergodic grain, as considered for example by \cite{De_Roeck_2017_1, Luitz_2017}. $ H_\om+h_\om$ is the total Hamiltonian and $ e^{-i(H_\om+h_\om)t} $ the corresponding dynamics. The support of $h_\om$ spreads into the system giving rise to the following L-R bound: 
\begin{align} \label{bound_single_slow}
 \underset{\omega}{\mathbb{E}}  \b| [e^{i(H_\om+h_\om)t} \, h_{\om}\, e^{-i(H_\om+h_\om)t},B] \b| \le    K \, h_{max} \, \|B\|  \, e^{-\frac{\dist(h,B)}{\xi} } \left( t^\beta + 8 \,h_{max} \,  \frac{t^{\beta+1}}{\beta+1} \right) 
\end{align}  
The interpretation of the upper bound in \eqref{bound_single_slow} is straightforward: considering for simplicity that the term proportional to $ t^{\beta+1}$ dominates on $ t^{\beta}$, within the time $ c\,t_{max}^{\beta+1}=e^{\frac{\dist(h,B)}{2\xi}}$, then the RHS of  \eqref{bound_single_slow} is exponentially small if $ \dist(h,B) \ll \xi $. With $ \beta=1 $, the presence of the perturbation $ h_\om $ makes $ t_{max} $ of the order of the square root of the same time interval in the unperturbed case, but still exponentially long in the distance among the supports of $ h_\om $ and $ B $. In this sense we call the slow dynamics with $ \beta > 0 $ of equation \eqref{LR_bound_MBL} stable: the lightcone remains logarithmic.

In the context of the $ XY $ model with random magnetic field, that is equivalent, through a Jordan-Wigner transformation, to the Anderson model, the bound  \eqref{LR_bound_MBL} (with a $\sup_t $ taken before averaging) has been shown in \cite{Sims_Stolz_2012} to hold with $ \beta = 0 $, meaning that at any $ t $ there is no spread at all of the support of $ A $. It should be stressed that for disordered systems it is essential to average over the different realizations of the Hamiltonian corresponding to different disorder configurations. This is crucial already at the level of the single particle Anderson model \cite{Stolz_2011,Simon_1995}.  When the dynamics of $ H_\om $ is of Anderson type, namely with $ f(t) = \textrm{constant} $ in \eqref{bound_comm_t}, theorem \ref{0} agrees with previous results on the destabilization of the many-body Anderson phase by perturbations, considered in \cite{De_Roeck_2020,Nach_2021,Huang_2023}.

Our theorem \ref{0} can be generalized, with slightly more restrictive assumptions, to systems in dimensions larger than one, we give a sketch of such generalization after the end of the proof of \ref{0}. A hint towards the existence of an MBL phase in dimensions larger than one comes by the experiments \cite{Bordia_2017,Choi_2016} and theoretical work like \cite{Wahl_2019}. If such a phase would be characterized by a L-R bound like in \eqref{LR_bound_MBL} then our theorem \ref{0} implies the stability of its dynamics.

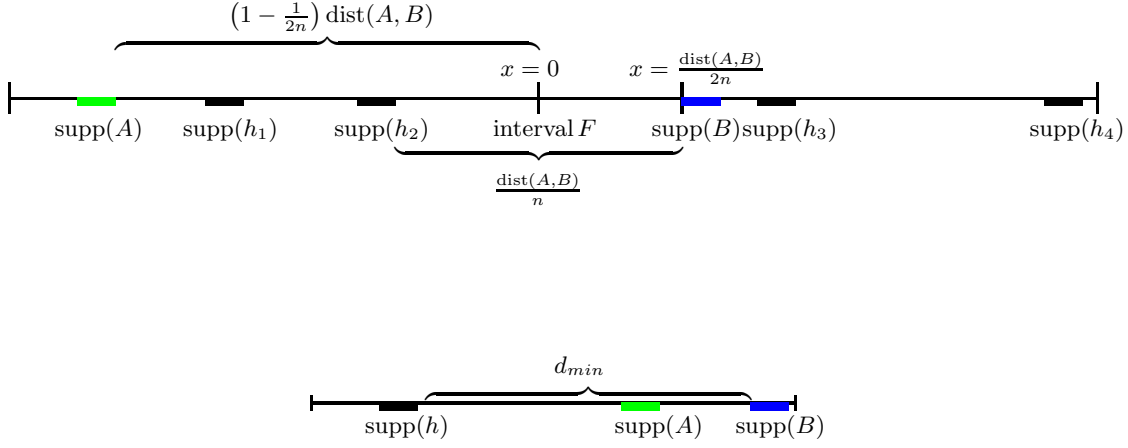
\begin{figure}[h]
\setlength{\unitlength}{1mm} 
\begin{picture}(140,40)(-30,0)

\thicklines

\put(-39,16){\line(0,1){4}}
\put(31,16){\line(0,1){4}}
\put(50,16){\line(0,1){4}}


\put(25,18){\line(-1,0){64}}

\put(50,18){\line(-1,0){64}}

\linethickness{1mm}
{\color{green}\put(-25,17.5){\line(-1,0){5}}}

\put(-8,17.5){\line(-1,0){5}}

\put(12,17.5){\line(-1,0){5}}

\thinlines

\put(-33,13){$ \supp(A) $}
\put(-16,13){$ \supp(h_1) $}
\put(4,13){$ \supp(h_2) $}


\put(-25,24){$\overbrace{\hspace{56mm}}$}

\put(-10,28){$\left(1-\frac{1}{2n}\right)\dist(A,B)$}

\thicklines

\put(105,16){\line(0,1){4}}

\put(105,18){\line(-1,0){64}}

\linethickness{1mm}
{\color{blue}\put(55,17.5){\line(-1,0){5}}}

\put(65,17.5){\line(-1,0){5}}

\put(103,17.5){\line(-1,0){5}}

\thinlines

\put(46,13){$ \supp(B) $}
\put(58,13){$ \supp(h_3) $}
\put(96,13){$ \supp(h_4) $}

\put(12,12){$\underbrace{\hspace{38mm}}$}

\put(43,21){$x=\frac{\dist(A,B)}{2n}$}

\put(26,21){$x=0$}

\put(25,5){$\frac{\dist(A,B)}{n}$}
\put(25,13){$ \textrm{interval} \, F$}

\end{picture}

\setlength{\unitlength}{1mm} 
\begin{picture}(140,40)(10,0)

\thicklines

\put(41,17){\line(0,1){2}}
\put(105,17){\line(0,1){2}}

\put(105,18){\line(-1,0){64}}

\linethickness{1mm}
\put(55,17.5){\line(-1,0){5}}

{\color{green}\put(87,17.5){\line(-1,0){5}}}

{\color{blue}\put(103,17.5){\line(-1,0){5}}}

\thinlines

\put(47,14){$ \supp(h) $}
\put(80,14){$ \supp(A) $}
\put(96,14){$ \supp(B) $}

\put(55,18){$\overbrace{\hspace{43mm}}$}



\put(72,22){$d_{min}$}


\end{picture}
\caption{Sketching a possible configuration of supports of the operators $ A $, $ B $ and $ h_j $ of the lattice $ \Lambda $. Below a configuration with perturbations that are far away from $ A $ and $ B $, as described by \eqref{bound_comm_far}.} 
\label{in_out}
\end{figure}

We now proceed to state our main result, theorem \ref{0}. Next we will provide several comments and consider special cases that illustrate the meaning of it. We will also present the dual picture for the description of the dynamics that arises from theorem \ref{0}, as mentioned above. The essence of theorem \ref{0} is that for slow dynamics of the MBL-type, like in equation \eqref{LR_bound_MBL}, any local perturbation that leaves free a significant part of the interval in between $ A $ and $ B $, see figure \ref{in_out}, and other than that is arbitrary, even with exponentially large norm like in lemma \ref{1}, leaves invariant the slow dynamics of the system in between $ A $ and $ B $, as quantified by the following Lieb-Robinson bounds \eqref{bound_full_t}, \eqref{bound_comm_far} and \eqref{bound_single}. The proof of theorem \ref{0} isspreads into the system giving rise to the followingspreads into the system giving rise to the followingspreads into the system giving rise to the following given at the end of the paper.

In the following, with abuse of notation, we will denote $ \dist(A,B) $ the distance among the supports of the two operators, namely $ \dist(\textrm{supp}(A),\textrm{supp}(B)) $. 
We explicitly formulate theorem \ref{0} for Hamiltonians that depend on disorder, $ \omega $, or more in general that depend on a parameter. Our result applies also in the deterministic case.




\begin{Theorem} \label{0}
{\it A one dimensional spin system, defined on the lattice $ \Lambda $, local Hilbert space $ \mathds{C}^r $, and with Hamiltonian $ H_\omega + \sum_j h_{j,\omega}(t) $, where both $ H_\omega $ and $ \sum_j h_{j,\omega}(t) $ are local operators, is given. For all $ j $, $ h_{j,\omega}(t) $ is supported on simply connected regions, that we assume independent from time and the disorder configuration, and we define $ h_{max}(t) := \max_{j,\omega} \|h_{j,\omega}(t)\| < \infty$. Two operators $ A $ and $ B $ supported on the simply connected regions $ \textrm{supp}(A) $ and $ \textrm{supp}(B) $ are also given. The dynamics of $ H $ is assumed to give rise to the following L-R bound, meaning that for all pairs of operators $ W $ and $ Z $, with supports at a distance $ \dist(W,Z) $, it holds:
\begin{align} \label{bound_comm_t}
 \underset{\omega}{\mathbb{E}}  \b| [e^{iH_\omega t} W e^{-iH_\omega t},Z] \b| \le   \, K \, \|W\| \, \|Z\| \, e^{-\frac{\dist(W,Z)}{\xi} }  \,  f(t)
\end{align}
with $ K $ a constant of $ O(1) $ independent from the size of the supports of $ W $ and $ Z $, and $ f $ positive, non decreasing.  
The operator of unitary evolution of the full system $ H_\omega + \sum_j h_{j,\omega}(t) $, is given by:
\begin{align} \label{total_dyn}
 V_\om(t):= T \left[ \exp \left(-i \left( H_\omega t + \int_0^t ds \, \sum_j h_{j,\omega}(s) \right) \right) \right]
\end{align}
We assume that the largest interval free of perturbations $ h_{j,\omega}(t) $ in between the supports of $ A $ and $ B $ has size $ \frac{\dist(A,B)}{n} $, with $ n \ge 1 $, see figure \ref{in_out}. The index $j$ in $ h_{j,\omega}(t) $ denotes the distance of its support from the center of such interval. On the left of $ A $ and on the right of $ B $ there can be a generic configuration of perturbations $ h_{j,\omega} $.  Then, it holds:
\begin{align} \label{bound_full_t}
\underset{\omega}{\mathbb{E}}   \b| [V_\om^*(t) \, A \, V_\om(t),B] \b| \le   K \|A\| \, \|B\|  \, e^{-\frac{\dist(A,B)}{2n\xi} } \, f(t) + 16 \, K \, \|A\| \, \|B\| \, \xi \, e^{-\frac{\dist(A,B)}{2n\xi}} \int_0^t ds f(s) h_{max}(s) 
\end{align}
If all the perturbations $ h_{j,\omega} $ are such that $ \dist(h_{j,\omega},B) \ge \dist(A,B) $, then denoting $ d_{min} := \min_j \{ \dist(h_{j,\omega},B) \} $, under the assumption \eqref{bound_comm_t}, it holds: 
\begin{align} \label{bound_comm_far}
\underset{\omega}{\mathbb{E}}   \b| [V_\om^*(t) \, A \, V_\om(t),B] \b| \le   K \, \|A\| \, \|B\|  \, e^{-\frac{\dist(A,B)}{\xi} } \,  f(t) + 16 \, K \, \|A\| \, \|B\| \, \xi \, e^{-\frac{d_{min}}{\xi}} \int_0^t ds f(s) h_{max}(s) 
\end{align}
If there is only one perturbation $ h(t) $ in the system and we look at how its support spreads, again under the assumption \eqref{bound_comm_t}, it holds: 
\begin{align} \label{bound_single}
\underset{\omega}{\mathbb{E}}   \b| [V_\om^*(t) \, h_{\om}(t) \, V_\om(t),B] \b| \le    K \, h_{max} \, \|B\|  \, e^{-\frac{\dist(h,B)}{\xi} } \,  f(t) + 8 \, K \, h_{max}(t) \, \|B\| \, e^{-\frac{\dist(h,B)}{\xi} } \,   \int_0^t ds \, f(s)  \,h_{max}(s)
\end{align} } 
\end{Theorem}

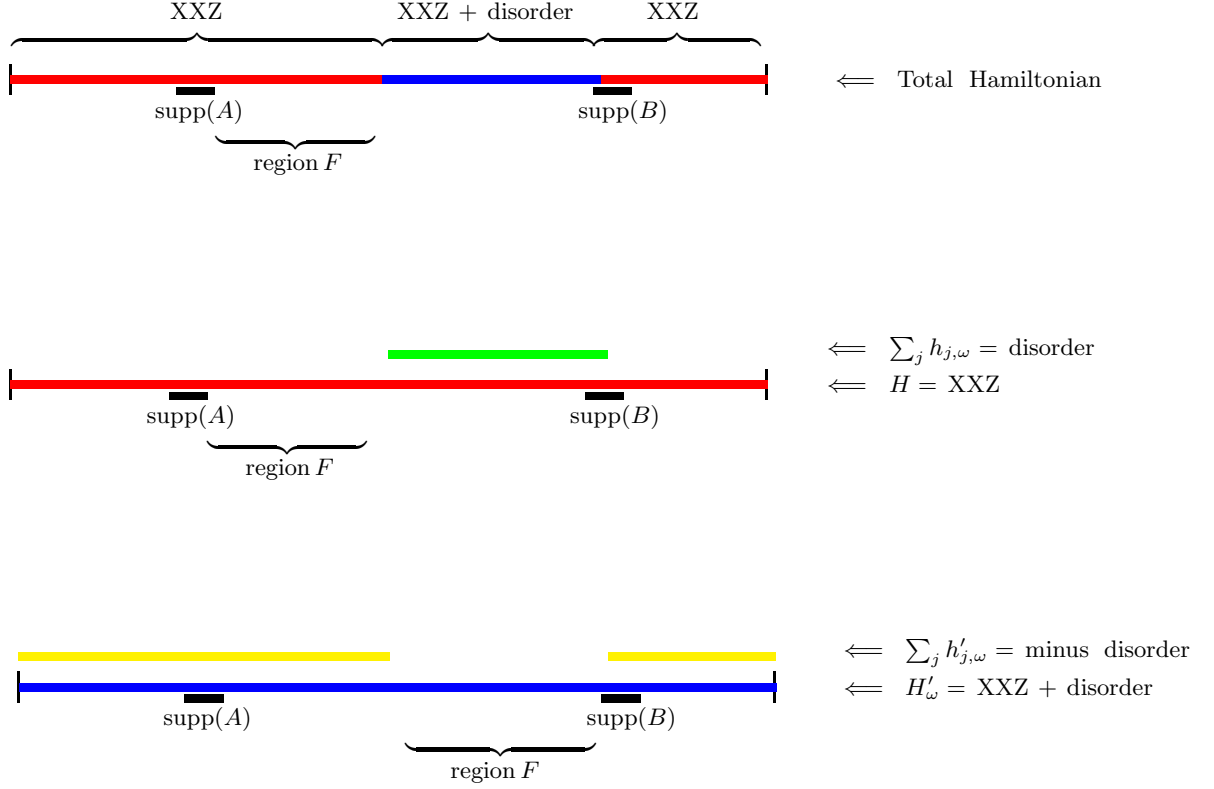
\begin{figure}[h]
\setlength{\unitlength}{1mm} 
\begin{picture}(180,50)(-30,0)

\thicklines

\put(-15,16){\line(0,1){4}}
\put(85,16){\line(0,1){4}}
\put(-15,18){\line(1,0){100}}

\linethickness{1mm}

\put(12,16.5){\line(-1,0){5}}
{\color{blue} \put(64,18){\line(-1,0){30}}}
{\color{red} \put(-16,18){\line(1,0){49}}}
{\color{red} \put(61,18){\line(1,0){22}}}
\put(65,16.5){\line(-1,0){5}}

\thinlines

\put(2,13){$ \supp(A) $}
\put(58,13){$ \supp(B) $}
\put(10,11){$\underbrace{\hspace{21mm}}$}
\put(15,6){$ \textrm{region} \, F$}

\put(4,26){$\textrm{XXZ}$}
\put(-17,22){$\overbrace{\hspace{49mm}}$}

\put(34,26){$\textrm{XXZ} \, +  \, \textrm{disorder} $}
\put(32,22){$\overbrace{\hspace{28mm}}$}

\put(67,26){$\textrm{XXZ} $}
\put(60,22){$\overbrace{\hspace{22mm}}$}

\put(92,17){$ \Longleftarrow \hspace{2mm} \textrm{Total} \hspace{2mm} \textrm{Hamiltonian} $}

\end{picture}


\begin{picture}(180,90)(-30,-50)

\thicklines

\put(-15,16){\line(0,1){4}}
\put(85,16){\line(0,1){4}}
\put(-15,18){\line(1,0){100}}

\linethickness{1mm}

{\color{green} \put(35,22){\line(1,0){29}}}

{\color{red} \put(-16,18){\line(1,0){100}}}

\put(65,16.5){\line(-1,0){5}}
\put(10,16.5){\line(-1,0){5}}

\thinlines

\put(92,17){$ \Longleftarrow \hspace{2mm} H= \,\textrm{XXZ} $}

\put(92,22){$ \Longleftarrow \hspace{2mm} \sum_j h_{j,\omega}=\,\textrm{disorder} $}

\put(2,13){$ \supp(A) $}
\put(58,13){$ \supp(B) $}
\put(10,11){$\underbrace{\hspace{21mm}}$}
\put(15,6){$ \textrm{region} \, F$}

\thicklines

\put(-15,-24){\line(0,1){4}}
\put(85,-24){\line(0,1){4}}
\put(-15,-22){\line(1,0){100}}

\linethickness{1mm}

\put(12,-23.5){\line(-1,0){5}}

{\color{yellow} \put(-15,-18){\line(1,0){49}}}
{\color{yellow} \put(62,-18){\line(1,0){22}}}

{\color{blue} \put(-17,-22){\line(1,0){100}}}
\put(65,-23.5){\line(-1,0){5}}

\thinlines

\put(2,-27){$ \supp(A) $}
\put(58,-27){$ \supp(B) $}
\put(34,-29){$\underbrace{\hspace{25mm}}$}
\put(40,-34){$ \textrm{region} \, F$}

\put(92,-23){$ \Longleftarrow \hspace{2mm} H_\omega'= \,\textrm{XXZ} \, + \, \textrm{disorder}$}

\put(92,-18){$ \Longleftarrow \hspace{2mm} \sum_j h_{j,\omega}' =\,\textrm{minus} \hspace{2mm} \textrm{disorder} $}

\end{picture}
\caption{A representation of the ``dual picture'' for the study of dynamics
from  Lieb-Robinson bounds. The final Hamiltonian, that is a regular $ XXZ$ model with a disordered region, corresponds to two different pairs $ (H,\sum_jh_j)$ and $ (H',\sum_j h_j')$. } 
\label{dual}
\end{figure}

%




In the following lemma we generalize the result of theorem \ref{0} allowing the perturbations $\{ h_{j,\omega}(t) \} $ to become very large far away from the supports of $ A $ and $ B $. It is physically sound that such perturbations shouldn't matter, within certain time scales, as far as the dynamics of distant regions is concerned. For simplicity we consider only the generalization of the upper bound \eqref{bound_comm_far}.

\begin{Lemma} \label{1}
 {\it  Let us consider the same assumptions as in theorem \ref{0} with the only difference of allowing even exponential growth of the perturbations away from $ x=0 $, (taken as the mid point among the supports of $ A $ and $ B $) as in figure \ref{in_out}, namely assuming that exist $ C (t) $ and $ \eta > \xi $ such that: $ \|h_{j,\omega}(t)\| \le C (t) \, e^{\frac{j}{\eta} }$. 
If all the perturbations $ \{h_{j,\omega}\} $ are such that $ \dist(h_{j,\omega},B) \ge \dist(A,B) $, then denoting $ d_{min} := \min_j \{ \dist(h_{j,\omega},B) \} $, and $ \frac{1}{\alpha}:= \frac{1}{\xi}-\frac{1}{\eta} > 0$, it holds: 
\begin{align} \label{bound_comm_far_extra}
\underset{\omega}{\mathbb{E}}   \b| [V_\om^*(t) \, A \, V_\om(t),B] \b| \le   2\, K \, \|A\| \, \|B\|  \, e^{-\frac{\dist(A,B)}{\xi} } \,  f(t) + 16 \, K \, \|A\| \, \|B\| \, \alpha \, e^{-\frac{d_{min}}{\alpha}} \int_0^t ds f(s) \, C(s) 
\end{align} }

\end{Lemma}

Before the proofs, that are presented in the secion \ref{sec_Methods}, we provide several remarks about the bounds \eqref{bound_full_t}, \eqref{bound_comm_far} and \eqref{bound_single}.

\section{Consequences of the stability of slow dynamics from theorem \ref{0}}

\subsection{Physical meaning of the assumptions in theorem \ref{0}}

The size of the interval $ F $ free from perturbations, as in figure \ref{in_out}, is $\frac{\dist(A,B)}{n}$. Despite theorem \ref{0} being mathematically correct for all $ n \ge 1 $, a meaningful physical interpretation requires $ n $ of order $ 1 $. The smaller the free region, the larger $ n $, the larger will become the influence of the perturbations on the L-R bound affecting the factor $ e^{-\frac{\dist(A,B)}{2n\xi}} $ in \eqref{bound_full_t}. The reason to ask for such interval free from perturbations, is that, for a fixed time $ t $ the meaning of locality is that increasing the distance among the supports of $ A $ and $ B $, the bound \eqref{bound_full_t} must decrease and become very small for large distances. Other than leaving an interval $ F $ free, there is no limit to the number of perturbations $\{ h_{j,\om} \}$ that are present in the system. They can have overlapping supports and be, for example, an actual nearest neighbour Hamiltonian $ \sum_k h_{k,k+1,\om} $. A proof independent from ours of the bound \eqref{bound_single} for the case of a single perturbation has also appeared in \cite{Capel_2024}, equation (55).

\subsection{Dual picture: consequences for fast dynamics}

We now take full advantage of the fact that the upper bound in \eqref{bound_full_t} is non perturbative and present a ``{\it dual picture}'' for the description of the dynamics that we mentioned before. We consider a pair of Hamiltonians $ H $ and $ H' $, and perturbations $ \sum_j h_j $, $ \sum_j h_j' $, such that $ H + \sum_j h_j = H' + \sum_j h_j' $. For example, let us fix: 
\begin{align} \label{HXXZ}
 H = \sum_{j=-L}^{L-1} \left( \sigma_j^x\sigma_{j+1}^x + \sigma_j^y\sigma_{j+1}^y + \Delta \sigma_j^z\sigma_{j+1}^z \right) 
\end{align}
with $\Delta \in (0,1]$, and the perturbation
\begin{align}
 \sum_{j=1}^R h_{j,\omega_j} = \sum_{j=1}^R \omega_j \sigma_j^z
\end{align}
providing on site disorder over the region $ R $, according to the figure \ref{dual}, with $ \omega_j $, for instance, taken at random uniformly from $ [-W,W] $. In this case $ \omega = \{\omega_j\}_{j\in \{1,...,R\}} $. The same total Hamiltonian is obtained, for every single realization of the disorder, namely of the variable $ \omega $, with
\begin{align} \label{do}
 H_\omega' = \sum_{j=-L}^{L-1} \left( \sigma_j^x\sigma_{j+1}^x + \sigma_j^y\sigma_{j+1}^y + \Delta \sigma_j^z\sigma_{j+1}^z  +  \omega_j \sigma_j^z \right)
\end{align}
and 
\begin{align} \label{undo}
 \sum_{j\in R^c} h_{j,\omega_j}' = - \sum_{j\in R^c} \omega_j \sigma_j^z 
\end{align}
where we have denoted $ R^c $ the complement of $ R $ in $ [-L,L]$.  In equation \eqref{undo} the variables $ \omega_j $ are the same as in \eqref{do}, therefore the terms $- \sum_{j\in R^c} \omega_j \sigma_j^z  $  ``undo'' the part of $ H' $ depending on disorder every where but on $ R $. Let us assume that $ H' $ gives rise to a slow dynamics, like in \eqref{LR_bound_MBL}, and we apply theorem \ref{0} to the Hamiltonian $ H' + \sum_j h_j' $. This implies that the support of $ A $, as in figure \ref{dual}, spreads slowly. In this way we have established that the region of disorder in proximity of the support of $ A $ crucially affects the time scale of the spread of the support of $ A $, that is logarithmic. Considering instead the dynamics arising from $ H $, equation \eqref{HXXZ}, with $ f(t) $, like in \eqref{bound_comm_t}, given by $ f(t)=e^{v\,t}$, we would have obtained that the time scale for the spread of the support of $ A $ was linear. The dual description of the dynamics allows a much tighter bound. A recent rigorous result hinting is this direction is given in section 4.5 of \cite{Gebert_2022}.  In the proof of theorem \ref{0} we show that taking $ h_j' $ depending on disorder, namely on $ \omega_j $, is consistent with taking averages over disorder configurations $\omega $.

\begin{figure}[h]
\setlength{\unitlength}{1mm} 

\begin{picture}(180,50)(-30,-50)

\thicklines

\put(-15,-24){\line(0,1){4}}
\put(85,-24){\line(0,1){4}}
\put(-15,-22){\line(1,0){100}}

\linethickness{1mm}

\put(40,-23.5){\line(-1,0){5}}
\put(82,-23.5){\line(-1,0){5}}


{\color{red} \put(-15,-18){\line(1,0){49}}}

{\color{blue} \put(-16,-22){\line(1,0){100}}}

\thinlines

\put(32,-27){$ \supp(A) $}
\put(74,-27){$ \supp(B) $}
\put(34,-29){$\underbrace{\hspace{50mm}}$}
\put(40,-34){$ H_\omega = \,\textrm{XXZ} \, + \, \textrm{disorder}$}
\put(48,-38){$ \textrm{Slow Dynamics}$}
\put(-16,-29){$\underbrace{\hspace{50mm}}$}
\put(0,-34){$ H=\textrm{XXZ}$}
\put(-3,-38){$ \textrm{Fast Dynamics}$}

\put(92,-23){$ \Longleftarrow \hspace{2mm} H_\omega = \,\textrm{XXZ} \, + \, \textrm{disorder}$}

\put(92,-18){$ \Longleftarrow \hspace{2mm} \sum_j h_{j,\omega} =\,\textrm{minus} \hspace{2mm} \textrm{disorder} $}

\end{picture}
\caption{As a consequence of \eqref{bound_comm_far} the dynamics on the right half of the spin chain remains slow.} 
\label{junction}
\end{figure}
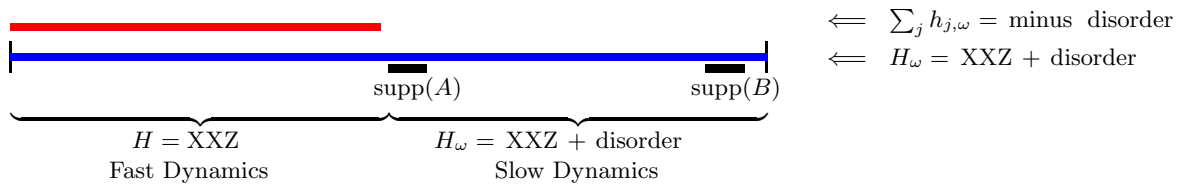

\subsection{Many body localization: avalanches spread exponentially slowly}

In the context of the study of the stability of many-body-localized phases the phenomenon of avalanches have gained importance starting with the work of De Roeck and Huveneers \cite{De_Roeck_2017_1, Luitz_2017}, and more recently, for example, in the works \cite{Morningstar_Huse_2022, Morningstar_Huse_2023}, where the effects of the presence of regions of anomalous low disorder are crucial in destabilizing the MBL phase.  An experimental signature of the avalanche scenario has been claimed in \cite{Leonard_2023}. Recent numerical results which agree with our result of avalanches spreading exponentially slowly are, for example, \cite{Sierant_2024_PRB,Scocco_2024}. 
When $ h_\om $ in \eqref{bound_single_slow} is a so called avalanche, as discussed in the introduction, the consequence of our bound \eqref{bound_single_slow} is that the avalanche spreads exponentially slowly. A recent work hinting towards MBL phases being more stable than thought is \cite{Goihl_2019}. 
More explicitly let us consider a perturbation $ h_\om $ that ``removes'' the Hamiltonian $ H $ on the region $[0,n] $, and replaces it, for example, with a random matrix. Considering the nearest neighbour Hamiltonian $ H_\om = \sum_{j=-L}^L H_{j,j+1,\om} $, this is realized by the perturbation $ h_\om = - \sum_{k=1}^n H_{k,k+1,\om} + T(n) $. The resulting Hamiltonian connects the random matrix $ T(n) $, for example in the GOE ensemble of size $ r^n \times r^n $ on the left with
$ \sum_{j=-L}^{-1} H_{j,j+1,\om} $ and on the right with $ \sum_{j=n}^{L} H_{j,j+1,\om} $. In  \cite{De_Roeck_2017_1,Luitz_2017} a similar setting was considered but allowing couplings of the random matrix with distant sites that decrease exponentially fast with the distance. The result of \eqref{bound_single_slow} is that the spread of $ e^{i(H_\om+h_\om)t}\,h_\om\,e^{-i(H_\om+h_\om)t} $ in time, starting from the region $ [0,n] $, happens according to a logarithmic light cone. In the work \cite{Potirniche_2019}  a GOE matrix has been modelled as an all to all Hubbard model. The bound \eqref{bound_single_slow} is meaningful within the time scale such that the right hand side of \eqref{bound_single_slow} is smaller than the trivial bound to the commutator, that is $ 2 \, h_{max} \, \|B\| $, therefore the larger the size of the GOE matrix, the larger its norm, the shorter the time scale that gives saturation of the bound.

 In a similar way we can consider the dynamics on one side of the junction among two systems as the result of a perturbation that affects half of the system. In this context it clearly appears that our theorem 1 and lemma 2 are non perturbative results. Let us consider the setting of figure \ref{junction}: we assume that the Hamiltonian $ H_\om $, for example as that in \eqref{do}, gives rise in equation \eqref{bound_comm_t} to a slow dynamics and take as a perturbation 
\begin{align} 
 \sum_{j=-1}^{-L} h_{j,\omega_j}' = - \sum_{j=-1}^{-L} \omega_j \sigma_j^z  \nonumber
\end{align} 
that cancels the disorder in the left half of the system. For any two operators $ A $ and $ B $ supported on the right half of the system the application of equation \eqref{bound_comm_far} shows that the dynamics is slow. Two recent works \cite{Sierant_2024_PRB, Scocco_2024} present numerical findings in agreement with this consequence of our theory.

\subsection{Dynamical instability of Anderson localization}

We now comment upon models with $f(t)=\textrm{constant}$ in \eqref{bound_comm_t}. The one dimensional $ XX $ or $ XY $ models with random magnetic field, that can be shown to be equivalent to the single particle Anderson model, have been studied for example by \cite{Sims_Stolz_2012}, section 4. Let us consider a perturbation $ h $ supported at one edge of the system. Our theorem \ref{0} gives predictions that qualitatively agree with the findings of \cite{Huang_2023}. The author of \cite{Huang_2023} has numerical evidences that the coupling of a $ XX $ model plus random magnetic field with a $ ZZ $ boundary term causes the entanglement-entropy dynamics to transition from uniformly bounded in time, like proven in \cite{Pastur_2014, Stolz_2015}, to $ \log t $, as it would be in a MBL system, \cite{Znidaric_2008,Bardarson_2012,Serbyn_2013}. Also in \cite{Gebert_2016} instabilities of a perturbed $ XY $ model leading to a dynamical growth of the L-R commutator were upper bounded. 
 The fact that the upper bound in \eqref{bound_single_slow} with $ \beta=0 $ goes like $ t $, would lead to the $ \log t $ growth of the entanglement entropy, as shown in \cite{Kim_2014,Zeng_2023}. Another way of quantifying the slow dynamics of MBL is employing quantum information related concepts like the Holevo quantity \cite{Nico_Katz_2022}. 
 The authors of \cite{De_Roeck_2020} and \cite{Nach_2021} have rigorously shown that in a model of exponential dynamical localization, that is $ \beta=0$ in \eqref{LR_bound_MBL}, perturbed by an Hamiltonian with a sparse distribution of interactions the transport is subdiffusive. Our result agrees with their, in fact the second term in the upper bound \eqref{bound_full_t} would grow like $ t $ in this case. Nevertheless within our theory we cannot rigorously conclude that exponential dynamical localization is destabilized by interactions, or by a boundary term, in fact for this purpose we would need a lower bound in \eqref{bound_full_t} rather than an upper bound. 
  In the context of the study of Schr\"{o}dinger operators Damanik and Tcheremchantsev \cite{Damanik_2008} have developed methods to evaluate lower bounds to the dynamics.

\subsection{Lieb-Robinson bounds and phases of matter}

A given Lieb-Robinson bound can give rise, in the context of systems with fast dynamics, to different types of transport regimes. The relation among the shape of the light cone as predicted by the Lieb-Robinson bound and the transport properties, both of particles and energy is not in a one to one correspondence. For example the authors of \cite{Kim_2013} have established that linear increase in time of entanglement entropy (that can be put in relation with a linear light cone \cite{Bravyi_Hastings_Verstraete_2006}) happens in a system with diffusive energy transport. Related work on entanglement spreading in diffusive systems has been done in \cite{Rakovszky_2019,Znidaric_2023}. Also in \cite{Rakovszky_2018} a linear lightcone, as predicted by the so called out-of-time-ordered-correlation function (OTOC), coexists with diffusive spreading of conserved charges in a quantum unitary circuit.  On the other hand a spin system with a linear lightcone like the $ XXZ $ model supports ballistic transport \cite{Prosen_2011}. This suggests that the details of the transport properties cannot be capture by the light cones structure \cite{Mahoney_2024}, that nevertheless can distinguish fast dynamics, leading to ballistic or diffusive transport, from slow dynamics leading to no transport of particles, like in MBL.

\subsection{Stability of slow dynamics for adiabatic and Floquet evolutions}

We consider two further particular cases related to \eqref{bound_full_t}, \eqref{bound_comm_far} and \eqref{bound_single}.

 The first regards the case of slow dynamics, like in \eqref{LR_bound_MBL}, arising from $ H $ and a single perturbation $ h(t) $  varying with a time scale $ \tau $: $ h(t) = h(\frac{t}{\tau}) $. Assuming that $ \|h(t)\| $ and  $ f(t) $  vary continuously in time, it is a consequence of the mean value theorem the existence of $ s' \in [0,\frac{t}{\tau}] $, such that:
\begin{align} \label{bound_full_adia}
   \int_0^t ds  \, \|h(s)\| \, f(s) = \|h(s')\| \, \int_0^t ds \, f(s) 
\end{align} 
This connects with what stated by the authors of \cite{Khemani_2015}. They have numerically found that a local perturbation, adiabatically varying with a time scale $ \tau $, is such that, in the infinite-time limit of the dynamics, $ t \rightarrow \infty $, and for an infinite system, it causes effects like charge redistribution over a length scale of the order of $ \log \tau $, both for the Anderson and the MBL dynamics. According to \cite{Khemani_2015} an arbitrarily slow rate of the adiabatic evolution would cause effects on an arbitrarily long length scale. Our upper bound \eqref{bound_single} in this case is such that for every finite time $ t \le \bar{t} $,  if $ \tau \gg \bar{t} $,  with $ h(0)=0 $ and $ \sup_{s'\in[0,\frac{t}{\tau}]} \|h(s')\| \ll \max \{\|A\|,\|B\|\}  $, then the effect of the perturbation $ h(t) $ is negligible. In fact despite propagating within a logarithmic light-cone, and therefore reaching a region of size at most of order $ \log \bar{t} $, the effect of the perturbation is suppressed by the small value of $ \|h(s')\| $ respect to the values of $ \|A\| $ and $ \|B\| $. If instead the initial value of the perturbation $ h(0) $ is of the order of $ \max \{\|A\|,\|B\|\} $ then the perturbation $ h(t) $ has a stronger effect but does not change the nature of the slow dynamics \eqref{LR_bound_MBL}.

The second case concerns perturbations varying periodically, $ h(t)=h(t+T) $, with high frequency.  An application of the Riemann-Lebesgue lemma implies the decoupling of the integral in \eqref{bound_single} as follows:
\begin{align} \label{bound_full_periodic}
   \int_0^t ds \, \|h(s)\| \, f(s)  = \frac{1}{T}\int_0^T ds \, \|h(s)\| \, \int_0^t ds \, f(s) + O\left(T \max_{s\in[0,t]} \|h(s)\| \max_{s\in[0,t]} f(s)  \right)
\end{align}
We see that in the limit of high frequency, or equivalently short period $ T $, such that 
\begin{align}
 O\left(T \max_{s\in[0,t]} \|h(s)\| \max_{s\in[0,t]} f(s)  \right) \ll 1
\end{align}
the integral in \eqref{bound_full_periodic} decouples in the product of two factors, one is the average of $ \|h(s)\| $ over a period, and the other one is the integral of $ f(s) $.

\subsection{Extension of theorem \ref{0} to higher dimensional systems}

\begin{figure}[h!]
\begin{tikzpicture}[scale=1]
\useasboundingbox (-4.5,-3) rectangle (20,4);
\thicklines

\draw[line width=1.5mm] (3.5,0) -- (4.5,0); 

\draw[blue, thick](4,0) circle (1);
\draw[green, thick](4,0) circle (1.5);
\draw[blue, thick](4,0) circle (2);

%

\filldraw[color=black, fill=black, thick](0.5,1.5) rectangle (0.3,1.3);
\node[black,above] at (0.4,0.7) {$h_1$};

\filldraw[color=black, fill=black, thick](4,0.5) rectangle (4.3,0.2);
\node[black,above] at (4.6,0.1) {$h_2$};

\filldraw[color=black, fill=black, thick](3.5,2.7) rectangle (3.8,3);
\node[black,above] at (3.7,2.2) {$h_3$};

\node[black,above] at (1.7,2.2) {$\textrm{Outer Region}$};

\filldraw[color=black, fill=black, thick](0,-0.1) rectangle (0.3,0.3);

 \node (test) at (4,0.7) {};
    \node (testDesc) [above right = of test] {Inner Region};
    \draw [->,thick] (testDesc) to [in = 60, out = -120] (test);

\node (test) at (5,-1) {};
    \node (testDesc) [below right = of test] {Supports Boundary};
    \draw [->,thick] (testDesc) to [in = -60, out = 120] (test);

\node[black,above] at (0.2,-0.7) {$\supp(A)$};
\node[black,above] at (4,-0.7) {$\supp(B)$};

\draw [decorate,thick,decoration={brace,amplitude=3pt,mirror,raise=4pt},xshift=-2pt,yshift=0pt]
(4.2,-1) -- (4.2,-2) node [black,midway,xshift=-5mm] 
{\footnotesize $ 2r $};
%

\end{tikzpicture}

\caption{A sketch in two dimensions of a configuration of perturbations that allows the generalization of \eqref{bound_full_t} to higher dimensions. The inner and outer regions are divided apart from a region free of perturbations, a circular corona in 2D, a spherical one in 3D.} 
\label{high_dim}
\end{figure}
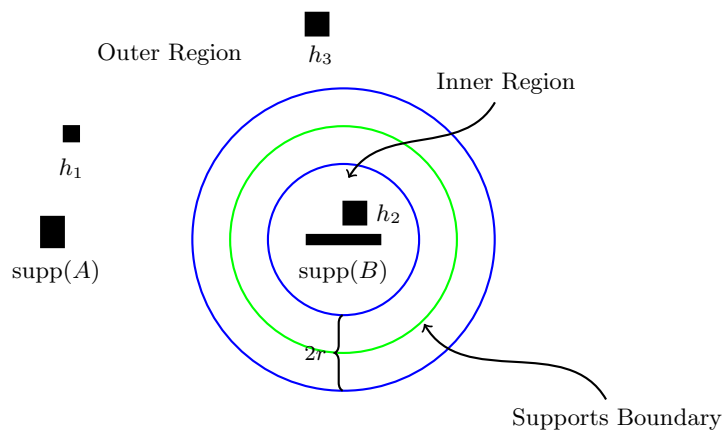

The extension of our theorem \ref{0} to systems of dimensionality $ d>1$ is straightforward under  slightly more stringent assumptions, that consist in requiring the existence of a corona around the support of $ B $ free of perturbations, as in figure \ref{high_dim}. This splits the space into an inner and outer regions. Clearly in the one dimensional case this is equivalent to say that not only there is a region free of perturbations on the left of $ B $, see figure \ref{in_out}, as asked by the assumption of theorem \ref{0}, but also on its right. The bound corresponding to \eqref{bound_full_t} in $ d>1$ would look substantially unaltered, but with a dependence of $ K $ on the dimensionality, see for instance \cite{Masanes_2009},  and  on $ \min \{| \partial \, \textrm{supp}(A)|,| \partial \, \textrm{supp}(B)| \} $ according to equation S17 of \cite{Yin_2023}, and with the factor $ e^{-\frac{\dist(A,B)}{2n\xi}} $ replaced by $ e^{-\frac{r}{\xi}} $ where we have denoted $ 2r $ the thickness of the corona free of perturbations, as in figure \ref{high_dim}. We give a sketch of the proof of these statements after the end of the proof of theorem \ref{0}.

 We also note that in \cite{Hastings_2008, Foss-Feig_2015, Kuwahara_2020, Baldwin_2023} already appeared the idea of combining the interaction picture and L-R bounds (see below the proof of theorem \ref{0}) but the contexts and the way in which this idea was applied are different from ours.

\section{Proofs} \label{sec_Methods}

The proofs of theorem \ref{0} and lemma \ref{1} follow.

\begin{proof}
To lighten the notation we drop the dependence of $ H $ and $h_j$ on $ \omega $, but we restore it when we want to stress averaging. 

The so called interaction picture (or representation) of quantum mechanics is based upon:
\begin{align} \label{int_picture}
V(t):= T \left[ \exp \left(-i \left( H t + \int_0^t ds \, h(s) \right) \right) \right] = e^{-iHt} \, T \left[ \exp \left(-i \int_0^t ds e^{iHs} \, h(s) \, e^{-iHs} \right) \right]
\end{align}
On the left hand side of \eqref{int_picture} there is the operator of time evolution of the full system with Hamiltonian $ H + h(t) $. $ T(\cdot) $ denotes time ordering, according to the rule: latest time goes on the left,  $ T^{-1} (\cdot) $ instead is the inverse time ordering: the latest time goes on the right, that arises by Hermitian conjugation. This allows to accordingly write the integrals arising from the exponential. Equation \eqref{int_picture} is proven showing that both operators are solutions of a differential equation with a unique solution (once the initial condition $V(0)=\mathds{1}$ is set), therefore they coincide.
\begin{align}
  & i \frac{d}{dt} T \left[ \exp \left(-i \left( H t + \int_0^t ds \, h(s) \right) \right) \right]  = (H+h(t)) T \left[ \exp \left(-i \left( H t + \int_0^t ds \, h(s) \right) \right) \right]  \\
  & i \frac{d}{dt} \left( e^{-iHt} \, T \left[ \exp \left(-i \int_0^t ds e^{iHs} \, h(s) \, e^{-iHs} \right) \right] \right) = \\
  & = H e^{-iHt} \, T \left[ \exp \left(-i \int_0^t ds e^{iHs} \, h(s) \, e^{-iHs} \right) \right] + e^{-iHt}  e^{iHt} \, h(t) \, e^{-iHt} \, T \left[ \exp \left(-i \int_0^t ds e^{iHs} \, h(s) \, e^{-iHs} \right) \right] \nonumber \\
  & = (H+h(t)) \, e^{-iHt} \, T \left[ \exp \left(-i \int_0^t ds e^{iHs} \, h(s) \, e^{-iHs} \right) \right] 
\end{align}

We now show that if two operators $ h_1(t) $ and $ h_2(t) $ are such that $ \forall (s_1,s_2) $, $ [h_1(s_1), h_2(s_2)]=0 $, then:
\begin{align} \label{prod_time}
 T \left( \exp \left(-i \int_0^t ds \, (h_1(s) + h_2(s))  \, \right)\right) = T \left( \exp \left(-i \int_0^t ds \, h_1(s) \, \right)\right) T \left( \exp \left(-i \int_0^t ds \, h_2(s) \, \right)\right) 
\end{align}
The argument is that since the LHS and the RHS of \eqref{prod_time} satisfy the same differential equation then they coincide. We have that:
\begin{align}
 & i \frac{d}{dt} \left[ T \left( \exp \left(-i \int_0^t ds \, h_1(s) \, \right)\right) T \left( \exp \left(-i \int_0^t ds \, h_2(s) \, \right)\right) \right] = \\
 & h_1(t) T \left( \exp \left(-i \int_0^t ds \, h_1(s) \, \right)\right) T \left( \exp \left(-i \int_0^t ds \, h_2(s) \, \right)\right) + \nonumber \\
 & + T \left( \exp \left(-i \int_0^t ds \, h_1(s) \, \right)\right) h_2(t) T \left( \exp \left(-i \int_0^t ds \, h_2(s) \, \right)\right) \\
 & = (h_1(t)+h_2(t)) T \left( \exp \left(-i \int_0^t ds \, h_1(s) \, \right)\right) T \left( \exp \left(-i \int_0^t ds \, h_2(s) \, \right)\right)
\end{align}
With $ V(t) $ as in \eqref{total_dyn} it is:
\begin{align} \label{comm_sum}
 \b| [V^*(t) \, A \, V(t),B] \b| = \b| [e^{iHt} A e^{-iHt}, T \left( e^{-i \int_0^t ds  \sum_j e^{-iHs}h_j(s) e^{iHs} } \right)  B  T^{-1}\left( e^{ i \int_0^t ds  \sum_j e^{-iHs} h_j(s) e^{iHs}} \right)] \b|
\end{align}
Let us define
\begin{align} 
\left(e^{iHt} A e^{-iHt}\right)_{B_a(A)}  := \frac{1}{r^{|B_a(A)^c|}} \left( \Tr_{B_a(A)^c} e^{iHt}(t) A e^{-iHt} \right) \otimes \mathds{1}_{B_a(A)^c} 
\end{align}
that is the partial trace on the Hilbert space associated to the complement, in $ \Lambda $, of the ball $ B_a(A) $, that identifies the union of $ \supp(A) $ with the region around $ \supp(A) $ within the distance $ a $. $\frac{1}{r^{|B_a(A)^c|}} $ is just a normalization factor, with $ |B_a(A)^c| $ denoting the number of sites (the cardinality) of the region $ B_a(A)^c $.

One of the main ideas in the following is to use the fact that the Lieb-Robinson bound \eqref{bound_comm_t} is equivalent to the bound:
\begin{align} \label{bound_restr_dyn}
   \b| e^{iHt} A e^{-iHt} - \left(e^{iHt} A e^{-iHt}\right)_{B_a(A)} \b| \le   \, K \, \|A\| \, f(t) \, e^{-\frac{a}{\xi} }
\end{align}
The fact that \eqref{bound_comm_t} implies \eqref{bound_restr_dyn} is shown using Haar-integration, as follows. 
With the support of $ A $ contained in $ X $, and $ V $ a unitary, it holds:
\begin{equation}
 \frac{1}{2^{|X^c|}} \left( \Tr_{X^c} V^* A V \right) \otimes \mathds{1}_{X^c} = \int_{W_{ \textrm{supported on}\, X^c}}  \, W^* (V^* A V) W d_{Haar}W \nonumber
\end{equation} 
The integral above is performed according to the Haar measure over all the unitaries $ W $ supported on the complement of $ X $, denoted $ X^c $. Then with $U:=U(t)$, it follows
\begin{align}
 & \| \frac{1}{2^{|X^c|}} \left( \Tr_{X^c} U^* A U \right) \otimes \mathds{1}_{X^c} - U^* A U \| = \| \int_{W \textrm{supported on}\, X^c}   \left( W^* (U^* A U) W - U^* A U \right)  d_{Haar}W   \| \nonumber \\
 & \le  \int_{W \textrm{supported on}\, X^c} \| [ U^* A U, W] \|  d_{Haar}W  \le K \|A\| f(t) e^{-\frac{\dist(\supp(A),X^c)}{\xi}} \nonumber
\end{align}

The fact that \eqref{bound_restr_dyn} implies \eqref{bound_comm_t} is easily seen as follows:
\begin{align}
& \b| [e^{iHt} A e^{-iHt},B] \b| = \b| [e^{iHt} A e^{-iHt} - \left(e^{iHt} A e^{-iHt}\right)_{B_{\dist(A,B)}(A)} + \left(e^{iHt} A e^{-iHt}\right)_{B_{\dist(A,B)(A)}} ,B] \b| \\ 
& \le 2 \|B\| \, \b| e^{iHt} A e^{-iHt} - \left(e^{iHt} A e^{-iHt}\right)_{B_{\dist(A,B)(A)}} \b| \le
\, K \, \|A\|\,\|B\| \, f(t) \, e^{-\frac{\dist(A,B)}{\xi} }
\end{align}
In the following definition of  $  (\widehat{h_j})_s $ and $ (\overline {h_j})_s $ to have a simple notation we assume the system's lattice to be $ \Lambda = [-L,L] \cap \mathds{Z} $.

We assume for simplicity that the center of the region $ F $ free from perturbations in figure \ref{in_out} corresponds to $ x=0 $. We restrict the supports of $ e^{-iHs} h_j(s) e^{iHs} $, with $ h_j(s) $ supported on the left of the region $ F $, to $ [-L,0] $ and call them $ (\widehat{h_j})_s $, while the perturbations $ h_j(s) $ supported on the right of $ F $ are such that $ e^{-iHs} h_j(s) e^{iHs} $ are replaced with the corresponding ones with support on the region $ [1,L] $. The more distant the support of $ h_j(s) $ from  $ F $, on the right or on the left, the smallest the error associated with the restriction of the support that we have just described.

\begin{align}
& (\widehat{h_j})_s := \frac{1}{r^{|[-L,0]^c|}} \left( \Tr_{[-L,0]^c} e^{-iHs} h_j(s) e^{iHs} \right) \otimes \mathds{1}_{[-L,0]^c} \hspace{5mm} \textrm{with} \hspace{2mm} \supp(h_j(s)) \hspace{2mm}  \textrm{on the left of $ F $} \label{hat} \\
& (\overline {h_j})_s := \frac{1}{r^{|[1,L]^c|}} \left( \Tr_{[1,L]^c} e^{-iHs} h_j(s) e^{iHs} \right) \otimes \mathds{1}_{[1,L]^c} \hspace{5mm} \textrm{with} \hspace{2mm}  \supp(h_j(s)) \hspace{2mm}   \textrm{on the right of $ F $} 
\end{align}
Each $ (\widehat{h_j})_s $ commutes with each $ (\overline {h_j})_s $ because their supports are disjoint. This implies that we can make use of equation \eqref{prod_time}. Defining 
\begin{align}
 \widehat{T} := T \left( e^{-i \int_0^t ds  \sum_j (\widehat{h_j})_s } \right) \hspace{5mm} \textrm{and} \hspace{5mm} 
 \overline{T} := T \left( e^{-i \int_0^t ds  \sum_k (\overline{h_k})_s } \right) \label{T_hat}
\end{align}
it holds:
\begin{align} \label{splitting}
T \left( e^{-i \int_0^t ds \left( \sum_j (\widehat{h_j})_s + \sum_k  (\overline {h_k})_s  \right) } \right) = \widehat{T} \, \overline{T} = \overline{T} \, \widehat{T} 
\end{align}
Following Appendix B of \cite{Tran_2019}, we now show how to upper bound in norm the difference among two time-ordered operators. With $ G(s) $ and $ E(s) $ norm-continuous Hermitian operators, it is:
\begin{align}
 &\b| T \left[ \exp \left(-i \int_0^t ds G(s) \right) \right] - T \left[ \exp \left(-i \int_0^t ds E(s) \right) \right] \b| \label{Duhamel_time_ordeer} \\
 &= \b| T^{-1} \left[ \exp \left(i \int_0^t ds G(s) \right) \right] T \left[ \exp \left(-i \int_0^t ds E(s) \right) \right] - \mathds{1} \b| \\
 &=\b| \int_0^t dr \, \frac{d}{dr} \left(
 T^{-1} \left[ \exp \left(i \int_0^r ds G(s) \right) \right] T \left[ \exp \left(-i \int_0^r ds E(s) \right) \right] \right) \b| \\
 &=\b| \int_0^t dr \left(
 T^{-1} \left[ \exp \left(i \int_0^r ds G(s) \right) \right]  i \left( G(r)-E(r) \right) T \left[ \exp \left(-i \int_0^r ds E(s) \right) \right] \right) \b| \\
 & \le \int_0^t dr \b| G(r)-E(r) \b|  
\end{align}
It follows from equation \eqref{Duhamel_time_ordeer} that:
\begin{align}
& \b| T \left( e^{-i \int_0^t ds \, \sum_j e^{-iHs} h_j(s) e^{iHs} } \right) - \widehat{T} \, \overline{T} \b| \label{first_term}  \\
& = \b| \int_0^t dr \, \frac{d}{dr} \left(
 T^{-1} \left[ \exp \left(i \int_0^r ds  \, \sum_j e^{-iHs} h_j(s) e^{iHs} \right) \right] \widehat{T} \, \overline{T} \right) \b| \\
& = \b| \int_0^t dr \, 
 T^{-1} \left[ \exp \left(i \int_0^r ds  \, \sum_j e^{-iHs} h_j(s) e^{iHs} \right) \right] i \left(\sum_j e^{-iHs} h_j(s) e^{iHs} - \sum_l (\widehat{h_l})_s - \sum_k (\overline{h_k})_s   \right) \widehat{T} \, \overline{T}  \b| \\ 
&\le \int_0^t ds \left( \sum_j \b| e^{-iHs} h_j(s) e^{iHs} - (\widehat{h_j})_s \b| +  \sum_k \b| e^{-iHs} h_k(s) e^{iHs} - (\overline{h_k})_s \b| \right) \label{last_term}
\end{align}  
To upper bound \eqref{last_term} we realize that the operator $ A $ in the Lieb-Robinson bound \eqref{bound_comm_t} can be explicitly time dependent and disorder dependent. Let us take $ C $ an operator with simply connected support and with $ \dist(C,h) = a $, then:
\begin{align}
 & \Eom \, \b| [e^{iH_\om t} h_\om(t) e^{-iH_\om t}, C] \b| = \Eom \b| [ h_\om(t) , e^{-iH_\om t} C e^{iH_\om t}] \b|  \\
 & = \Eom \, \b| [ h_\om(t) , e^{-iH_\om t} C e^{iH_\om t} -\left(e^{-iH_\om t} C e^{iH_\om t}\right)_{B_a(C)} ] \b| \\
  & \le \Eom \, \, 2 \b|  h_\om(t) \b| \b| e^{-iH_\om t} C e^{iH_\om t} -\left(e^{-iH_\om t} C e^{iH_\om t}\right)_{B_a(C)} \b| \\
 & \le 2 K \, h_{max}(t) \|C\| \, f(t) \, e^{-\frac{a}{\xi}} \label{LR_bound_dis}
\end{align}
Where we have introduced the maximal norm over all the local perturbations $ h_{i,\om} $, $ h_{max}(s) := \max_{i,\om} \{ \|h_{i,\om}(s)\| \} $. \eqref{LR_bound_dis} in turn implies, using Haar-integration as mentioned above, that:
\begin{align} \label{bound_restr_dyn_h}
 \Eom \,  \b| e^{iH_\om t} h_\om(t) e^{-iH_\om t} - \left(e^{iH_\om t} h_\om(t) e^{-iH_\om t}\right)_{B_a(h)} \b| \le 2 K \, h_{max}(t) \, f(t) \, e^{-\frac{a}{\xi} }
\end{align}
We can now go back to \eqref{last_term} considering the contribution to the integral in \eqref{last_term} of the $ h_j $ supported on the left of the interval $ F $ free from perturbations.

\begin{align} 
&\sum_j \b| e^{-iHs} h_j(s) e^{iHs} - (\widehat{h_j})_s \b| \le  2K \, f(s) \sum_j \|h_j(s)\|  \, e^{-\frac{\dist(\supp(h_j),x=0)}{\xi} } \label{sum_1} \\
& \le  2K \, f(s) \, h_{max}(s) \, \sum_j   \, e^{-\frac{\dist(\supp(h_j),x=0)}{\xi} }
 \le  2K \, f(s) \, h_{max}(s) \, \sum_{j=\frac{\dist(A,B)}{2n}+1}^\infty  e^{-\frac{j}{\xi} } \label{sum_2}
\end{align}
Observing that:
\begin{align}
 \sum_{j=\frac{\dist(A,B)}{2n}+1}^\infty  e^{-\frac{j}{\xi} } = e^{-\left(\frac{\dist(A,B)}{2n}+1\right)\frac{1}{\xi} }  \sum_{j=0}^\infty  e^{-\frac{j}{\xi} } = e^{-\frac{\dist(A,B)}{2n\xi} }  \frac{1}{e^{\frac{1}{\xi}}-1} \le \xi \,  e^{-\frac{\dist(A,B)}{2n\xi} }
\end{align}
By symmetry the contribution to \eqref{last_term} of the $ h_j $ supported on the right of the region $ F $  is upper bounded by the same quantity. Restoring the average over the parameter $\om$, it turns out that \eqref{first_term} is upper bounded by:
\begin{align}
& \Eom \b| T \left( e^{-i \int_0^t ds \, \sum_j e^{-iH_\om s} h_{j,\om}(s) e^{iH_\om s} } \right) - \widehat{T} \, \overline{T} \b| \le  4 K \, \xi \,  e^{-\frac{\dist(A,B)}{2n\xi} } \int_0^t ds \, f(s) \, h_{max}(s)
\end{align}
To shorten the notation we define:
\begin{align} \label{T_op}
 T:= T \left( e^{-i \int_0^t ds \, \sum_j e^{-iHs} h_j(s) e^{iHs} } \right) 
\end{align}
Going back to \eqref{first_term} we have:

\begin{align}
& \b| V(t)^* \, A \, V(t) ,B] \b| = \b| \left[e^{iHt} A e^{-iHt},T \, B \, T^{-1}]\right] \b| \\
& =\b| [e^{iHt} A e^{-iHt} ,\left( T - \widehat{T} \, \overline{T} +  \widehat{T} \, \overline{T} \right) B \left( T^{-1} - \overline{T}^{-1} \, \widehat{T}^{-1}  - \overline{T}^{-1} \, \widehat{T}^{-1}  \right) ] \b| \\
& = \b| [e^{iHt} A e^{-iHt} ,\left( T - \widehat{T} \, \overline{T} \right) B  T^{-1} +
\widehat{T} \, \overline{T}   B \left( T^{-1} - \overline{T}^{-1} \, \widehat{T}^{-1} \right)  
 + \widehat{T} \, \overline{T} \,  B \overline{T}^{-1} \, \widehat{T}^{-1} ] \b| \label{make_T_commute}\\
& \le 4 \|A\| \, \|B\| \, \| T - \widehat{T} \, \overline{T} \| + 
\b| [e^{iHt} A e^{-iHt} , \overline{T} \, B \, \overline{T}^{-1} ] \b| \label{almost}
\end{align}
In the last steep we made use of the fact that $  \widehat{T} $ commutes with both $ B $ and $ \overline{T} $ because their supports are disjoint. Since $ \overline{T} \, B \, \overline{T}^{-1} $ is supported on $ x \ge 0 $, see figure \ref{in_out}, we apply the Lieb-Robinson bound \eqref{bound_comm_t} taking into account that the distance among the supports of $A$ and $\overline{T} \, B \, \overline{T}^{-1}$ equals $(1-\frac{1}{2n}) \dist(A,B) $, for the case depicted in \ref{in_out}. In general the shortest possible distance among these supports is $\frac{\dist(A,B)}{2n} $, corresponding to the case in which the interval free of perturbations $ F $ is contiguous to the support of $ A $. 
Restarting from \eqref{almost} we have:

\begin{align}
& \Eom \, \b| V(t)^* \, A \, V(t) ,B] \b| \\
& \le 4 \, \|A\| \, \|B\| \, \Eom \, \| T - \widehat{T} \, \overline{T} \| +
\Eom \, \b| [e^{iHt} A e^{-iHt}, \overline{T} \, B \, \overline{T}^{-1}  ] \b| \nonumber \\
& \le  K \, \|A\| \, \|B\| \,e^{-\frac{\dist(A,B)}{2n\xi} } + 16 \, K \, \|A\| \, \|B\| \, \xi \,  e^{-\frac{\dist(A,B)}{2n\xi} } \int_0^t ds \, f(s) \, h_{max}(s) 
\end{align}
This completes the proof of the bound given in \eqref{bound_full_t}.

We now sketch the proof of \eqref{bound_comm_far}, all the technical tools and concepts required for it have been already introduced above. We explicitly present the proof of \eqref{bound_comm_far} because the assumption that  $ \forall j $, $ \dist(h_{j,\omega},B) \ge \dist(A,B) $ is slightly more restrictive than being the region in between $ A $ and $ B $ free of perturbations. In fact setting $ n=1 $ in \eqref{bound_full_t} we get a factor $ \frac{1}{2} $ at the exponent of $ e^{-\frac{\dist(A,B)}{2\xi}} $, that is absent in \eqref{bound_comm_far} where we get $ e^{-\frac{\dist(A,B)}{2\xi}} $. Looking, in this setting, at figure \ref{in_out}, we see that the error introduced restricting the support of $ e^{iHt}h_je^{-iHt}$ on the left of the support of $ B $ or on its right is at most proportional to $ e^{-\frac{d_{min} }{\xi}} $. The summation of the contribution of all $ h_j $ is then performed as in \eqref{sum_1} and \eqref{sum_2}.

We end the proof briefly commenting on \eqref{bound_single}. 
If there is only one perturbation $ h(t) $, and in the L-R bound it is $A=h(t)$, the support of $ h(t) $ is on the left of that of $ B $, in this scenario we can apply the bound \eqref{bound_comm_far}. But with only one perturbation there is no summation, like in \eqref{sum_1}, \eqref{sum_2}, to be performed, therefore no factor $ \xi $ appearing together with $ h_{max}$ in the second term in the right hand side of \eqref{bound_single}.

This completes the proof of theorem \ref{0}.

\end{proof}

We provide a brief account for the generalization of the bound \eqref{bound_full_t} to higher dimensions. The main idea follows those presented in the proof of theorem  \ref{0} given above. Namely to split the time-ordered part of the operator of time evolution into two factors, following the logic of equation \eqref{splitting}. Referring to figure \ref{high_dim}, all the perturbations $ h_j $ with support in the outer region, are such that the support of $ e^{iHt}h_je^{-iHt}$ is restricted to the region outside the green circle (in three dimensions that would be a sphere) with an error of at most (for those perturbations sitting just outside the circular corona) of the order of $ e^{-\frac{r}{\xi}} $. The perturbations supported in the inner region instead, get the support of $ e^{iHt}h_je^{-iHt}$ restricted till the green circle, with an error      
again at most of the order of $ e^{-\frac{r}{\xi}} $. These two types of restricted operators give rise to a new pair $ \widehat{T} $ and $ \overline{T} $ that commutes as in \eqref{splitting}. The rest of the reasoning follows that of the proof of theorem \ref{0}.

Proof of lemma \ref{1}: The upper bound in \eqref{bound_comm_far_extra} follows immediately from inserting in \eqref{sum_1} the assumption $ \|h_{j,\omega}(t)\| \le C (t) \, e^{\frac{j}{\eta} }$.


%
%

\subsection*{Acknowledgments}
The authous  acknowledge support from UKRI grant EP/R029075/1. They're also glad to acknowledge discussions on the topic of Lieb-Robinson bounds and Many-Body-Localization with Abolfazl Bayat, Igor Lerner and George McArdle.


\bibliography{bibliography_Stability_Slow_Dynamics}

\begin{thebibliography}{62}%
\makeatletter
\providecommand \@ifxundefined [1]{%
 \@ifx{#1\undefined}
}%
\providecommand \@ifnum [1]{%
 \ifnum #1\expandafter \@firstoftwo
 \else \expandafter \@secondoftwo
 \fi
}%
\providecommand \@ifx [1]{%
 \ifx #1\expandafter \@firstoftwo
 \else \expandafter \@secondoftwo
 \fi
}%
\providecommand \natexlab [1]{#1}%
\providecommand \enquote  [1]{``#1''}%
\providecommand \bibnamefont  [1]{#1}%
\providecommand \bibfnamefont [1]{#1}%
\providecommand \citenamefont [1]{#1}%
\providecommand \href@noop [0]{\@secondoftwo}%
\providecommand \href [0]{\begingroup \@sanitize@url \@href}%
\providecommand \@href[1]{\@@startlink{#1}\@@href}%
\providecommand \@@href[1]{\endgroup#1\@@endlink}%
\providecommand \@sanitize@url [0]{\catcode `\\12\catcode `\$12\catcode
  `\&12\catcode `\#12\catcode `\^12\catcode `\_12\catcode `\%12\relax}%
\providecommand \@@startlink[1]{}%
\providecommand \@@endlink[0]{}%
\providecommand \url  [0]{\begingroup\@sanitize@url \@url }%
\providecommand \@url [1]{\endgroup\@href {#1}{\urlprefix }}%
\providecommand \urlprefix  [0]{URL }%
\providecommand \Eprint [0]{\href }%
\providecommand \doibase [0]{https://doi.org/}%
\providecommand \selectlanguage [0]{\@gobble}%
\providecommand \bibinfo  [0]{\@secondoftwo}%
\providecommand \bibfield  [0]{\@secondoftwo}%
\providecommand \translation [1]{[#1]}%
\providecommand \BibitemOpen [0]{}%
\providecommand \bibitemStop [0]{}%
\providecommand \bibitemNoStop [0]{.\EOS\space}%
\providecommand \EOS [0]{\spacefactor3000\relax}%
\providecommand \BibitemShut  [1]{\csname bibitem#1\endcsname}%
\let\auto@bib@innerbib\@empty
\bibitem [{\citenamefont {Lieb}\ and\ \citenamefont
  {Robinson}(1972)}]{Lieb_Robinson}%
  \BibitemOpen
  \bibfield  {author} {\bibinfo {author} {\bibfnamefont {E.~H.}\ \bibnamefont
  {Lieb}}\ and\ \bibinfo {author} {\bibfnamefont {D.}~\bibnamefont
  {Robinson}},\ }\bibfield  {title} {\bibinfo {title} {The finite group
  velocity of quantum spin systems},\ }\href
  {https://doi.org/10.1007/bf01645779} {\bibfield  {journal} {\bibinfo
  {journal} {Commun. Math. Phys.}\ }\textbf {\bibinfo {volume} {28}},\ \bibinfo
  {pages} {251} (\bibinfo {year} {1972})}\BibitemShut {NoStop}%
\bibitem [{\citenamefont {Hastings}(2004)}]{Hastings_2004}%
  \BibitemOpen
  \bibfield  {author} {\bibinfo {author} {\bibfnamefont {M.~B.}\ \bibnamefont
  {Hastings}},\ }\bibfield  {title} {\bibinfo {title} {Lieb-{Schultz}-{Mattis}
  in higher dimensions},\ }\href {https://doi.org/10.1103/PhysRevB.69.104431}
  {\bibfield  {journal} {\bibinfo  {journal} {Phys. Rev. B}\ }\textbf {\bibinfo
  {volume} {69}},\ \bibinfo {pages} {104431} (\bibinfo {year} {2004})},\
  \Eprint {https://arxiv.org/abs/cond-mat/0305505} {arXiv:cond-mat/0305505
  [cond-mat.str-el]} \BibitemShut {NoStop}%
\bibitem [{\citenamefont {Hastings}\ and\ \citenamefont
  {Michalakis}(2014)}]{Hastings_2014}%
  \BibitemOpen
  \bibfield  {author} {\bibinfo {author} {\bibfnamefont {M.~B.}\ \bibnamefont
  {Hastings}}\ and\ \bibinfo {author} {\bibfnamefont {S.}~\bibnamefont
  {Michalakis}},\ }\bibfield  {title} {\bibinfo {title} {Quantization of {Hall}
  conductance for interacting electrons on a torus},\ }\bibfield  {journal}
  {\bibinfo  {journal} {Communications in Mathematical Physics}\ }\textbf
  {\bibinfo {volume} {334}},\ \href {https://doi.org/10.1007/s00220-014-2167-x}
  {10.1007/s00220-014-2167-x} (\bibinfo {year} {2014}),\ \Eprint
  {https://arxiv.org/abs/1306.1258} {arXiv:1306.1258 [quant-ph]} \BibitemShut
  {NoStop}%
\bibitem [{\citenamefont {Bravyi}\ \emph {et~al.}(2006)\citenamefont {Bravyi},
  \citenamefont {Hastings},\ and\ \citenamefont
  {Verstraete}}]{Bravyi_Hastings_Verstraete_2006}%
  \BibitemOpen
  \bibfield  {author} {\bibinfo {author} {\bibfnamefont {S.}~\bibnamefont
  {Bravyi}}, \bibinfo {author} {\bibfnamefont {M.~B.}\ \bibnamefont
  {Hastings}},\ and\ \bibinfo {author} {\bibfnamefont {F.}~\bibnamefont
  {Verstraete}},\ }\bibfield  {title} {\bibinfo {title} {Lieb-{Robinson} bounds
  and the generation of correlations and topological quantum order},\ }\href
  {https://doi.org/10.1103/PhysRevLett.97.050401} {\bibfield  {journal}
  {\bibinfo  {journal} {Phys. Rev. Lett.}\ }\textbf {\bibinfo {volume} {97}},\
  \bibinfo {pages} {050401} (\bibinfo {year} {2006})},\ \Eprint
  {https://arxiv.org/abs/quant-ph/0603121} {arXiv:quant-ph/0603121}
  \BibitemShut {NoStop}%
\bibitem [{\citenamefont {Osborne}(2007)}]{Osborne_2007}%
  \BibitemOpen
  \bibfield  {author} {\bibinfo {author} {\bibfnamefont {T.~J.}\ \bibnamefont
  {Osborne}},\ }\bibfield  {title} {\bibinfo {title} {Simulating adiabatic
  evolution of gapped spin systems},\ }\href
  {https://doi.org/10.1103/PhysRevA.75.032321} {\bibfield  {journal} {\bibinfo
  {journal} {Phys. Rev. A}\ }\textbf {\bibinfo {volume} {75}},\ \bibinfo
  {pages} {032321} (\bibinfo {year} {2007})},\ \Eprint
  {https://arxiv.org/abs/quant-ph/0601019} {arXiv:quant-ph/0601019}
  \BibitemShut {NoStop}%
\bibitem [{\citenamefont {Gong}\ and\ \citenamefont
  {Hamazaki}(2022)}]{Gong_2022}%
  \BibitemOpen
  \bibfield  {author} {\bibinfo {author} {\bibfnamefont {Z.}~\bibnamefont
  {Gong}}\ and\ \bibinfo {author} {\bibfnamefont {R.}~\bibnamefont
  {Hamazaki}},\ }\bibfield  {title} {\bibinfo {title} {Bounds in nonequilibrium
  quantum dynamics},\ }\bibfield  {journal} {\bibinfo  {journal} {International
  Journal of Modern Physics B}\ }\textbf {\bibinfo {volume} {36}},\ \href
  {https://doi.org/10.1142/s0217979222300079} {10.1142/s0217979222300079}
  (\bibinfo {year} {2022}),\ \Eprint {https://arxiv.org/abs/2202.02011}
  {arXiv:2202.02011 [quant-ph]} \BibitemShut {NoStop}%
\bibitem [{\citenamefont {(Anthony)~Chen}\ \emph {et~al.}(2023)\citenamefont
  {(Anthony)~Chen}, \citenamefont {Lucas},\ and\ \citenamefont
  {Yin}}]{Lucas_review}%
  \BibitemOpen
  \bibfield  {author} {\bibinfo {author} {\bibfnamefont {C.-F.}\ \bibnamefont
  {(Anthony)~Chen}}, \bibinfo {author} {\bibfnamefont {A.}~\bibnamefont
  {Lucas}},\ and\ \bibinfo {author} {\bibfnamefont {C.}~\bibnamefont {Yin}},\
  }\bibfield  {title} {\bibinfo {title} {Speed limits and locality in many-body
  quantum dynamics},\ }\href {https://doi.org/10.1088/1361-6633/acfaae}
  {\bibfield  {journal} {\bibinfo  {journal} {Reports on Progress in Physics}\
  }\textbf {\bibinfo {volume} {86}},\ \bibinfo {pages} {116001} (\bibinfo
  {year} {2023})},\ \Eprint {https://arxiv.org/abs/2303.07386}
  {arXiv:2303.07386 [quant-ph]} \BibitemShut {NoStop}%
\bibitem [{\citenamefont {Nachtergaele}\ \emph {et~al.}(2019)\citenamefont
  {Nachtergaele}, \citenamefont {Sims},\ and\ \citenamefont
  {Young}}]{Nach_2019}%
  \BibitemOpen
  \bibfield  {author} {\bibinfo {author} {\bibfnamefont {B.}~\bibnamefont
  {Nachtergaele}}, \bibinfo {author} {\bibfnamefont {R.}~\bibnamefont {Sims}},\
  and\ \bibinfo {author} {\bibfnamefont {A.}~\bibnamefont {Young}},\ }\bibfield
   {title} {\bibinfo {title} {{Quasi-locality bounds for quantum lattice
  systems. I. {Lieb}-{Robinson} bounds, quasi-local maps, and spectral flow
  automorphisms}},\ }\href {https://doi.org/10.1063/1.5095769} {\bibfield
  {journal} {\bibinfo  {journal} {Journal of Mathematical Physics}\ }\textbf
  {\bibinfo {volume} {60}},\ \bibinfo {pages} {061101} (\bibinfo {year}
  {2019})},\ \Eprint {https://arxiv.org/abs/1810.02428} {arXiv:1810.02428
  [math-ph]} \BibitemShut {NoStop}%
\bibitem [{\citenamefont {Basko}\ \emph {et~al.}(2006)\citenamefont {Basko},
  \citenamefont {Aleiner},\ and\ \citenamefont {Altshuler}}]{Basko_2006}%
  \BibitemOpen
  \bibfield  {author} {\bibinfo {author} {\bibfnamefont {D.}~\bibnamefont
  {Basko}}, \bibinfo {author} {\bibfnamefont {I.}~\bibnamefont {Aleiner}},\
  and\ \bibinfo {author} {\bibfnamefont {B.}~\bibnamefont {Altshuler}},\
  }\bibfield  {title} {\bibinfo {title} {Metal–insulator transition in a
  weakly interacting many-electron system with localized single-particle
  states},\ }\href {https://doi.org/10.1016/j.aop.2005.11.014} {\bibfield
  {journal} {\bibinfo  {journal} {Annals of Physics}\ }\textbf {\bibinfo
  {volume} {321}},\ \bibinfo {pages} {1126–1205} (\bibinfo {year} {2006})},\
  \Eprint {https://arxiv.org/abs/cond-mat/0506617} {arXiv:cond-mat/0506617
  [math-ph]} \BibitemShut {NoStop}%
\bibitem [{\citenamefont {Gornyi}\ \emph {et~al.}(2005)\citenamefont {Gornyi},
  \citenamefont {Mirlin},\ and\ \citenamefont {Polyakov}}]{Gornyi_2005}%
  \BibitemOpen
  \bibfield  {author} {\bibinfo {author} {\bibfnamefont {I.~V.}\ \bibnamefont
  {Gornyi}}, \bibinfo {author} {\bibfnamefont {A.~D.}\ \bibnamefont {Mirlin}},\
  and\ \bibinfo {author} {\bibfnamefont {D.~G.}\ \bibnamefont {Polyakov}},\
  }\bibfield  {title} {\bibinfo {title} {Interacting electrons in disordered
  wires: {Anderson} localization and low-$t$ transport},\ }\href
  {https://doi.org/10.1103/PhysRevLett.95.206603} {\bibfield  {journal}
  {\bibinfo  {journal} {Phys. Rev. Lett.}\ }\textbf {\bibinfo {volume} {95}},\
  \bibinfo {pages} {206603} (\bibinfo {year} {2005})},\ \Eprint
  {https://arxiv.org/abs/cond-mat/0506411} {arXiv:cond-mat/0506411 [math-ph]}
  \BibitemShut {NoStop}%
\bibitem [{\citenamefont {Oganesyan}\ and\ \citenamefont
  {Huse}(2007)}]{Oganesyan_2007}%
  \BibitemOpen
  \bibfield  {author} {\bibinfo {author} {\bibfnamefont {V.}~\bibnamefont
  {Oganesyan}}\ and\ \bibinfo {author} {\bibfnamefont {D.~A.}\ \bibnamefont
  {Huse}},\ }\bibfield  {title} {\bibinfo {title} {Localization of interacting
  fermions at high temperature},\ }\href
  {https://doi.org/10.1103/PhysRevB.75.155111} {\bibfield  {journal} {\bibinfo
  {journal} {Phys. Rev. B}\ }\textbf {\bibinfo {volume} {75}},\ \bibinfo
  {pages} {155111} (\bibinfo {year} {2007})},\ \Eprint
  {https://arxiv.org/abs/cond-mat/0610854} {arXiv:cond-mat/0610854
  [cond-mat.str-el]} \BibitemShut {NoStop}%
\bibitem [{\citenamefont {$\check{Z}$nidari$\check{c}$}\ \emph
  {et~al.}(2008)\citenamefont {$\check{Z}$nidari$\check{c}$}, \citenamefont
  {Prosen},\ and\ \citenamefont {Prelov$\check{s}$ek}}]{Znidaric_2008}%
  \BibitemOpen
  \bibfield  {author} {\bibinfo {author} {\bibfnamefont {M.}~\bibnamefont
  {$\check{Z}$nidari$\check{c}$}}, \bibinfo {author} {\bibfnamefont
  {T.}~\bibnamefont {Prosen}},\ and\ \bibinfo {author} {\bibfnamefont
  {P.}~\bibnamefont {Prelov$\check{s}$ek}},\ }\bibfield  {title} {\bibinfo
  {title} {Many-body localization in the heisenberg {$XXZ$} magnet in a random
  field},\ }\href {https://doi.org/10.1103/PhysRevB.77.064426} {\bibfield
  {journal} {\bibinfo  {journal} {Phys. Rev. B}\ }\textbf {\bibinfo {volume}
  {77}},\ \bibinfo {pages} {064426} (\bibinfo {year} {2008})},\ \Eprint
  {https://arxiv.org/abs/0706.2539} {arXiv:0706.2539 [quant-ph]} \BibitemShut
  {NoStop}%
\bibitem [{\citenamefont {Abanin}\ \emph {et~al.}(2019)\citenamefont {Abanin},
  \citenamefont {Altman}, \citenamefont {Bloch},\ and\ \citenamefont
  {Serbyn}}]{Abanin_2019}%
  \BibitemOpen
  \bibfield  {author} {\bibinfo {author} {\bibfnamefont {D.~A.}\ \bibnamefont
  {Abanin}}, \bibinfo {author} {\bibfnamefont {E.}~\bibnamefont {Altman}},
  \bibinfo {author} {\bibfnamefont {I.}~\bibnamefont {Bloch}},\ and\ \bibinfo
  {author} {\bibfnamefont {M.}~\bibnamefont {Serbyn}},\ }\bibfield  {title}
  {\bibinfo {title} {Colloquium: Many-body localization, thermalization, and
  entanglement},\ }\href {https://doi.org/10.1103/RevModPhys.91.021001}
  {\bibfield  {journal} {\bibinfo  {journal} {Rev. Mod. Phys.}\ }\textbf
  {\bibinfo {volume} {91}},\ \bibinfo {pages} {021001} (\bibinfo {year}
  {2019})},\ \Eprint {https://arxiv.org/abs/1804.11065} {arXiv:1804.11065
  [cond-mat.dis-nn]} \BibitemShut {NoStop}%
\bibitem [{\citenamefont {Sierant}\ \emph {et~al.}(2024)\citenamefont
  {Sierant}, \citenamefont {Lewenstein}, \citenamefont {Scardicchio},
  \citenamefont {Vidmar},\ and\ \citenamefont {Zakrzewski}}]{Sierant_2024}%
  \BibitemOpen
  \bibfield  {author} {\bibinfo {author} {\bibfnamefont {P.}~\bibnamefont
  {Sierant}}, \bibinfo {author} {\bibfnamefont {M.}~\bibnamefont {Lewenstein}},
  \bibinfo {author} {\bibfnamefont {A.}~\bibnamefont {Scardicchio}}, \bibinfo
  {author} {\bibfnamefont {L.}~\bibnamefont {Vidmar}},\ and\ \bibinfo {author}
  {\bibfnamefont {J.}~\bibnamefont {Zakrzewski}},\ }\href@noop {} {\bibinfo
  {title} {Many-body localization in the age of classical computing}} (\bibinfo
  {year} {2024}),\ \Eprint {https://arxiv.org/abs/2403.07111} {arXiv:2403.07111
  [cond-mat.dis-nn]} \BibitemShut {NoStop}%
\bibitem [{\citenamefont {De~Roeck}\ and\ \citenamefont
  {Huveneers}(2017)}]{De_Roeck_2017_1}%
  \BibitemOpen
  \bibfield  {author} {\bibinfo {author} {\bibfnamefont {W.}~\bibnamefont
  {De~Roeck}}\ and\ \bibinfo {author} {\bibfnamefont {F.}~\bibnamefont
  {Huveneers}},\ }\bibfield  {title} {\bibinfo {title} {Stability and
  instability towards delocalization in many-body localization systems},\
  }\href {https://doi.org/10.1103/PhysRevB.95.155129} {\bibfield  {journal}
  {\bibinfo  {journal} {Phys. Rev. B}\ }\textbf {\bibinfo {volume} {95}},\
  \bibinfo {pages} {155129} (\bibinfo {year} {2017})},\ \Eprint
  {https://arxiv.org/abs/1608.01815} {arXiv:1608.01815 [cond-mat.dis-nn]}
  \BibitemShut {NoStop}%
\bibitem [{\citenamefont {Luitz}\ \emph {et~al.}(2017)\citenamefont {Luitz},
  \citenamefont {Huveneers},\ and\ \citenamefont {De~Roeck}}]{Luitz_2017}%
  \BibitemOpen
  \bibfield  {author} {\bibinfo {author} {\bibfnamefont {D.~J.}\ \bibnamefont
  {Luitz}}, \bibinfo {author} {\bibfnamefont {F.}~\bibnamefont {Huveneers}},\
  and\ \bibinfo {author} {\bibfnamefont {W.}~\bibnamefont {De~Roeck}},\
  }\bibfield  {title} {\bibinfo {title} {How a small quantum bath can
  thermalize long localized chains},\ }\href
  {https://doi.org/10.1103/PhysRevLett.119.150602} {\bibfield  {journal}
  {\bibinfo  {journal} {Phys. Rev. Lett.}\ }\textbf {\bibinfo {volume} {119}},\
  \bibinfo {pages} {150602} (\bibinfo {year} {2017})},\ \Eprint
  {https://arxiv.org/abs/1705.10807} {arXiv:1705.10807 [cond-mat.str-el]}
  \BibitemShut {NoStop}%
\bibitem [{\citenamefont {Kim}\ \emph {et~al.}(2014)\citenamefont {Kim},
  \citenamefont {Chandran},\ and\ \citenamefont {Abanin}}]{Kim_2014}%
  \BibitemOpen
  \bibfield  {author} {\bibinfo {author} {\bibfnamefont {I.~H.}\ \bibnamefont
  {Kim}}, \bibinfo {author} {\bibfnamefont {A.}~\bibnamefont {Chandran}},\ and\
  \bibinfo {author} {\bibfnamefont {D.~A.}\ \bibnamefont {Abanin}},\
  }\href@noop {} {\bibinfo {title} {Local integrals of motion and the
  logarithmic lightcone in many-body localized systems}} (\bibinfo {year}
  {2014}),\ \Eprint {https://arxiv.org/abs/1412.3073} {arXiv:1412.3073
  [cond-mat.dis-nn]} \BibitemShut {NoStop}%
\bibitem [{\citenamefont {Nachtergaele}\ and\ \citenamefont
  {Reschke}(2021)}]{Nach_2021}%
  \BibitemOpen
  \bibfield  {author} {\bibinfo {author} {\bibfnamefont {B.}~\bibnamefont
  {Nachtergaele}}\ and\ \bibinfo {author} {\bibfnamefont {J.}~\bibnamefont
  {Reschke}},\ }\bibfield  {title} {\bibinfo {title} {Slow propagation in some
  disordered quantum spin chains},\ }\href
  {https://doi.org/10.1007/s10955-020-02681-2} {\bibfield  {journal} {\bibinfo
  {journal} {Journal of Statistical Physics}\ }\textbf {\bibinfo {volume}
  {182}},\ \bibinfo {pages} {12} (\bibinfo {year} {2021})},\ \Eprint
  {https://arxiv.org/abs/1906.10167} {arXiv:1906.10167 [math-ph]} \BibitemShut
  {NoStop}%
\bibitem [{\citenamefont {Zeng}\ \emph {et~al.}(2023)\citenamefont {Zeng},
  \citenamefont {Hamma}, \citenamefont {Zhang}, \citenamefont {Liu},
  \citenamefont {Li}, \citenamefont {Fan},\ and\ \citenamefont
  {Liu}}]{Zeng_2023}%
  \BibitemOpen
  \bibfield  {author} {\bibinfo {author} {\bibfnamefont {Y.}~\bibnamefont
  {Zeng}}, \bibinfo {author} {\bibfnamefont {A.}~\bibnamefont {Hamma}},
  \bibinfo {author} {\bibfnamefont {Y.-R.}\ \bibnamefont {Zhang}}, \bibinfo
  {author} {\bibfnamefont {Q.}~\bibnamefont {Liu}}, \bibinfo {author}
  {\bibfnamefont {R.}~\bibnamefont {Li}}, \bibinfo {author} {\bibfnamefont
  {H.}~\bibnamefont {Fan}},\ and\ \bibinfo {author} {\bibfnamefont {W.-M.}\
  \bibnamefont {Liu}},\ }\href@noop {} {\bibinfo {title} {Logarithmic light
  cone, slow entanglement growth and scrambling, and quantum memory}} (\bibinfo
  {year} {2023}),\ \Eprint {https://arxiv.org/abs/2305.08334} {arXiv:2305.08334
  [quant-ph]} \BibitemShut {NoStop}%
\bibitem [{\citenamefont {Toniolo}\ and\ \citenamefont
  {Bose}(2024)}]{Toniolo_2024_2}%
  \BibitemOpen
  \bibfield  {author} {\bibinfo {author} {\bibfnamefont {D.}~\bibnamefont
  {Toniolo}}\ and\ \bibinfo {author} {\bibfnamefont {S.}~\bibnamefont {Bose}},\
  }\href@noop {} {\bibinfo {title} {Dynamics of many-body localized systems:
  {Lieb}-{Robinson} bounds and {Rényi} entropy}} (\bibinfo {year}
  {2024})\BibitemShut {NoStop}%
\bibitem [{\citenamefont {Elgart}\ and\ \citenamefont
  {Klein}(2024)}]{Elgart_2023}%
  \BibitemOpen
  \bibfield  {author} {\bibinfo {author} {\bibfnamefont {A.}~\bibnamefont
  {Elgart}}\ and\ \bibinfo {author} {\bibfnamefont {A.}~\bibnamefont {Klein}},\
  }\bibfield  {title} {\bibinfo {title} {Slow propagation of information on the
  random xxz quantum spin chain},\ }\href
  {https://doi.org/10.1007/s00220-024-05127-y} {\bibfield  {journal} {\bibinfo
  {journal} {Commun. Math. Phys.}\ }\textbf {\bibinfo {volume} {405}},\
  \bibinfo {pages} {239} (\bibinfo {year} {2024})},\ \Eprint
  {https://arxiv.org/abs/2311.14188} {arXiv:2311.14188 [math-ph]} \BibitemShut
  {NoStop}%
\bibitem [{\citenamefont {Smith}\ \emph {et~al.}(2017)\citenamefont {Smith},
  \citenamefont {Knolle}, \citenamefont {Kovrizhin},\ and\ \citenamefont
  {Moessner}}]{Smith_2017}%
  \BibitemOpen
  \bibfield  {author} {\bibinfo {author} {\bibfnamefont {A.}~\bibnamefont
  {Smith}}, \bibinfo {author} {\bibfnamefont {J.}~\bibnamefont {Knolle}},
  \bibinfo {author} {\bibfnamefont {D.~L.}\ \bibnamefont {Kovrizhin}},\ and\
  \bibinfo {author} {\bibfnamefont {R.}~\bibnamefont {Moessner}},\ }\bibfield
  {title} {\bibinfo {title} {Disorder-free localization},\ }\href
  {https://doi.org/10.1103/PhysRevLett.118.266601} {\bibfield  {journal}
  {\bibinfo  {journal} {Phys. Rev. Lett.}\ }\textbf {\bibinfo {volume} {118}},\
  \bibinfo {pages} {266601} (\bibinfo {year} {2017})},\ \Eprint
  {https://arxiv.org/abs/1701.04748} {arXiv:1701.04748 [cond-mat.str-el]}
  \BibitemShut {NoStop}%
\bibitem [{\citenamefont {Smith}\ \emph {et~al.}(2019)\citenamefont {Smith},
  \citenamefont {Knolle}, \citenamefont {Moessner},\ and\ \citenamefont
  {Kovrizhin}}]{Smith_2019}%
  \BibitemOpen
  \bibfield  {author} {\bibinfo {author} {\bibfnamefont {A.}~\bibnamefont
  {Smith}}, \bibinfo {author} {\bibfnamefont {J.}~\bibnamefont {Knolle}},
  \bibinfo {author} {\bibfnamefont {R.}~\bibnamefont {Moessner}},\ and\
  \bibinfo {author} {\bibfnamefont {D.~L.}\ \bibnamefont {Kovrizhin}},\
  }\bibfield  {title} {\bibinfo {title} {Logarithmic spreading of
  out-of-time-ordered correlators without many-body localization},\ }\href
  {https://doi.org/10.1103/PhysRevLett.123.086602} {\bibfield  {journal}
  {\bibinfo  {journal} {Phys. Rev. Lett.}\ }\textbf {\bibinfo {volume} {123}},\
  \bibinfo {pages} {086602} (\bibinfo {year} {2019})},\ \Eprint
  {https://arxiv.org/abs/1812.07981} {arXiv:1812.07981 [cond-mat.str-el]}
  \BibitemShut {NoStop}%
\bibitem [{\citenamefont {Yin}\ \emph {et~al.}(2024)\citenamefont {Yin},
  \citenamefont {Nandkishore},\ and\ \citenamefont {Lucas}}]{Yin_2024}%
  \BibitemOpen
  \bibfield  {author} {\bibinfo {author} {\bibfnamefont {C.}~\bibnamefont
  {Yin}}, \bibinfo {author} {\bibfnamefont {R.}~\bibnamefont {Nandkishore}},\
  and\ \bibinfo {author} {\bibfnamefont {A.}~\bibnamefont {Lucas}},\ }\bibfield
   {title} {\bibinfo {title} {Eigenstate localization in a many-body quantum
  system},\ }\href {https://doi.org/10.1103/PhysRevLett.133.137101} {\bibfield
  {journal} {\bibinfo  {journal} {Phys. Rev. Lett.}\ }\textbf {\bibinfo
  {volume} {133}},\ \bibinfo {pages} {137101} (\bibinfo {year} {2024})},\
  \Eprint {https://arxiv.org/abs/2405.12279} {arXiv:2405.12279
  [cond-mat.stat-mech]} \BibitemShut {NoStop}%
\bibitem [{\citenamefont {Hamza}\ \emph {et~al.}(2012)\citenamefont {Hamza},
  \citenamefont {Sims},\ and\ \citenamefont {Stolz}}]{Sims_Stolz_2012}%
  \BibitemOpen
  \bibfield  {author} {\bibinfo {author} {\bibfnamefont {E.}~\bibnamefont
  {Hamza}}, \bibinfo {author} {\bibfnamefont {R.}~\bibnamefont {Sims}},\ and\
  \bibinfo {author} {\bibfnamefont {G.}~\bibnamefont {Stolz}},\ }\bibfield
  {title} {\bibinfo {title} {Dynamical localization in disordered quantum spin
  systems},\ }\href {https://doi.org/10.1007/s00220-012-1544-6} {\bibfield
  {journal} {\bibinfo  {journal} {Commun. Math. Phys.}\ }\textbf {\bibinfo
  {volume} {315}},\ \bibinfo {pages} {215} (\bibinfo {year} {2012})},\ \Eprint
  {https://arxiv.org/abs/1108.3811} {arXiv:1108.3811 [math-ph]} \BibitemShut
  {NoStop}%
\bibitem [{\citenamefont {Stolz}(2011)}]{Stolz_2011}%
  \BibitemOpen
  \bibfield  {author} {\bibinfo {author} {\bibfnamefont {G.}~\bibnamefont
  {Stolz}},\ }\bibfield  {title} {\bibinfo {title} {An introduction to the
  mathematics of {Anderson} localization},\ }\href@noop {} {\bibfield
  {journal} {\bibinfo  {journal} {Contemp Math}\ }\textbf {\bibinfo {volume}
  {552}},\ \bibinfo {pages} {71} (\bibinfo {year} {2011})},\ \Eprint
  {https://arxiv.org/abs/1104.2317} {arXiv:1104.2317 [math-ph]} \BibitemShut
  {NoStop}%
\bibitem [{\citenamefont {del Rio}\ \emph {et~al.}(1995)\citenamefont {del
  Rio}, \citenamefont {Jitomirskaya}, \citenamefont {Last},\ and\ \citenamefont
  {Simon}}]{Simon_1995}%
  \BibitemOpen
  \bibfield  {author} {\bibinfo {author} {\bibfnamefont {R.}~\bibnamefont {del
  Rio}}, \bibinfo {author} {\bibfnamefont {S.}~\bibnamefont {Jitomirskaya}},
  \bibinfo {author} {\bibfnamefont {Y.}~\bibnamefont {Last}},\ and\ \bibinfo
  {author} {\bibfnamefont {B.}~\bibnamefont {Simon}},\ }\bibfield  {title}
  {\bibinfo {title} {What is localization?},\ }\href
  {https://doi.org/10.1103/PhysRevLett.75.117} {\bibfield  {journal} {\bibinfo
  {journal} {Phys. Rev. Lett.}\ }\textbf {\bibinfo {volume} {75}},\ \bibinfo
  {pages} {117} (\bibinfo {year} {1995})}\BibitemShut {NoStop}%
\bibitem [{\citenamefont {De~Roeck}\ \emph {et~al.}(2020)\citenamefont
  {De~Roeck}, \citenamefont {Huveneers},\ and\ \citenamefont
  {Olla}}]{De_Roeck_2020}%
  \BibitemOpen
  \bibfield  {author} {\bibinfo {author} {\bibfnamefont {W.}~\bibnamefont
  {De~Roeck}}, \bibinfo {author} {\bibfnamefont {F.}~\bibnamefont
  {Huveneers}},\ and\ \bibinfo {author} {\bibfnamefont {S.}~\bibnamefont
  {Olla}},\ }\bibfield  {title} {\bibinfo {title} {Subdiffusion in
  one-dimensional hamiltonian chains with sparse interactions},\ }\href
  {https://doi.org/10.1007/s10955-020-02496-1} {\bibfield  {journal} {\bibinfo
  {journal} {Journal of Statistical Physics}\ }\textbf {\bibinfo {volume}
  {180}},\ \bibinfo {pages} {678–698} (\bibinfo {year} {2020})},\ \Eprint
  {https://arxiv.org/abs/1909.07322v2} {arXiv:1909.07322v2 [math-ph]}
  \BibitemShut {NoStop}%
\bibitem [{\citenamefont {Huang}(2023)}]{Huang_2023}%
  \BibitemOpen
  \bibfield  {author} {\bibinfo {author} {\bibfnamefont {Y.}~\bibnamefont
  {Huang}},\ }\bibfield  {title} {\bibinfo {title} {Adding boundary terms to
  {Anderson} localized hamiltonians leads to unbounded growth of
  entanglement},\ }\href {https://doi.org/10.1209/0295-5075/acc19d} {\bibfield
  {journal} {\bibinfo  {journal} {Europhysics Letters}\ }\textbf {\bibinfo
  {volume} {142}},\ \bibinfo {pages} {10001} (\bibinfo {year} {2023})},\
  \Eprint {https://arxiv.org/abs/2109.07640} {arXiv:2109.07640
  [cond-mat.dis-nn]} \BibitemShut {NoStop}%
\bibitem [{\citenamefont {Bordia}\ \emph {et~al.}(2017)\citenamefont {Bordia},
  \citenamefont {L\"uschen}, \citenamefont {Scherg}, \citenamefont
  {Gopalakrishnan}, \citenamefont {Knap}, \citenamefont {Schneider},\ and\
  \citenamefont {Bloch}}]{Bordia_2017}%
  \BibitemOpen
  \bibfield  {author} {\bibinfo {author} {\bibfnamefont {P.}~\bibnamefont
  {Bordia}}, \bibinfo {author} {\bibfnamefont {H.}~\bibnamefont {L\"uschen}},
  \bibinfo {author} {\bibfnamefont {S.}~\bibnamefont {Scherg}}, \bibinfo
  {author} {\bibfnamefont {S.}~\bibnamefont {Gopalakrishnan}}, \bibinfo
  {author} {\bibfnamefont {M.}~\bibnamefont {Knap}}, \bibinfo {author}
  {\bibfnamefont {U.}~\bibnamefont {Schneider}},\ and\ \bibinfo {author}
  {\bibfnamefont {I.}~\bibnamefont {Bloch}},\ }\bibfield  {title} {\bibinfo
  {title} {Probing slow relaxation and many-body localization in
  two-dimensional quasiperiodic systems},\ }\href
  {https://doi.org/10.1103/PhysRevX.7.041047} {\bibfield  {journal} {\bibinfo
  {journal} {Phys. Rev. X}\ }\textbf {\bibinfo {volume} {7}},\ \bibinfo {pages}
  {041047} (\bibinfo {year} {2017})},\ \Eprint
  {https://arxiv.org/abs/1704.03063} {arXiv:1704.03063 [cond-mat.quant-gas]}
  \BibitemShut {NoStop}%
\bibitem [{\citenamefont {Choi}\ \emph {et~al.}(2016)\citenamefont {Choi},
  \citenamefont {Hild}, \citenamefont {Zeiher}, \citenamefont {Schauß},
  \citenamefont {Rubio-Abadal}, \citenamefont {Yefsah}, \citenamefont
  {Khemani}, \citenamefont {Huse}, \citenamefont {Bloch},\ and\ \citenamefont
  {Gross}}]{Choi_2016}%
  \BibitemOpen
  \bibfield  {author} {\bibinfo {author} {\bibfnamefont {J.-y.}\ \bibnamefont
  {Choi}}, \bibinfo {author} {\bibfnamefont {S.}~\bibnamefont {Hild}}, \bibinfo
  {author} {\bibfnamefont {J.}~\bibnamefont {Zeiher}}, \bibinfo {author}
  {\bibfnamefont {P.}~\bibnamefont {Schauß}}, \bibinfo {author} {\bibfnamefont
  {A.}~\bibnamefont {Rubio-Abadal}}, \bibinfo {author} {\bibfnamefont
  {T.}~\bibnamefont {Yefsah}}, \bibinfo {author} {\bibfnamefont
  {V.}~\bibnamefont {Khemani}}, \bibinfo {author} {\bibfnamefont {D.~A.}\
  \bibnamefont {Huse}}, \bibinfo {author} {\bibfnamefont {I.}~\bibnamefont
  {Bloch}},\ and\ \bibinfo {author} {\bibfnamefont {C.}~\bibnamefont {Gross}},\
  }\bibfield  {title} {\bibinfo {title} {Exploring the many-body localization
  transition in two dimensions},\ }\href
  {https://doi.org/10.1126/science.aaf8834} {\bibfield  {journal} {\bibinfo
  {journal} {Science}\ }\textbf {\bibinfo {volume} {352}},\ \bibinfo {pages}
  {1547–1552} (\bibinfo {year} {2016})},\ \Eprint
  {https://arxiv.org/abs/1604.04178} {arXiv:1604.04178 [cond-mat.quant-gas]}
  \BibitemShut {NoStop}%
\bibitem [{\citenamefont {Wahl}\ \emph {et~al.}(2018)\citenamefont {Wahl},
  \citenamefont {Pal},\ and\ \citenamefont {Simon}}]{Wahl_2019}%
  \BibitemOpen
  \bibfield  {author} {\bibinfo {author} {\bibfnamefont {T.~B.}\ \bibnamefont
  {Wahl}}, \bibinfo {author} {\bibfnamefont {A.}~\bibnamefont {Pal}},\ and\
  \bibinfo {author} {\bibfnamefont {S.~H.}\ \bibnamefont {Simon}},\ }\bibfield
  {title} {\bibinfo {title} {Signatures of the many-body localized regime in
  two dimensions},\ }\href {https://doi.org/10.1038/s41567-018-0339-x}
  {\bibfield  {journal} {\bibinfo  {journal} {Nature Physics}\ }\textbf
  {\bibinfo {volume} {15}},\ \bibinfo {pages} {164–169} (\bibinfo {year}
  {2018})},\ \Eprint {https://arxiv.org/abs/1711.02678} {arXiv:1711.02678
  [cond-mat.dis-nn]} \BibitemShut {NoStop}%
\bibitem [{\citenamefont {Ángela Capel}\ \emph {et~al.}(2024)\citenamefont
  {Ángela Capel}, \citenamefont {Moscolari}, \citenamefont {Teufel},\ and\
  \citenamefont {Wessel}}]{Capel_2024}%
  \BibitemOpen
  \bibfield  {author} {\bibinfo {author} {\bibnamefont {Ángela Capel}},
  \bibinfo {author} {\bibfnamefont {M.}~\bibnamefont {Moscolari}}, \bibinfo
  {author} {\bibfnamefont {S.}~\bibnamefont {Teufel}},\ and\ \bibinfo {author}
  {\bibfnamefont {T.}~\bibnamefont {Wessel}},\ }\href
  {https://arxiv.org/abs/2310.09182} {\bibinfo {title} {From decay of
  correlations to locality and stability of the gibbs state}} (\bibinfo {year}
  {2024}),\ \Eprint {https://arxiv.org/abs/2310.09182v2} {arXiv:2310.09182v2
  [math-ph]} \BibitemShut {NoStop}%
\bibitem [{\citenamefont {Gebert}\ \emph {et~al.}(2022)\citenamefont {Gebert},
  \citenamefont {Moon},\ and\ \citenamefont {Nachtergaele}}]{Gebert_2022}%
  \BibitemOpen
  \bibfield  {author} {\bibinfo {author} {\bibfnamefont {M.}~\bibnamefont
  {Gebert}}, \bibinfo {author} {\bibfnamefont {A.}~\bibnamefont {Moon}},\ and\
  \bibinfo {author} {\bibfnamefont {B.}~\bibnamefont {Nachtergaele}},\
  }\bibfield  {title} {\bibinfo {title} {A {Lieb}–{Robinson} bound for
  quantum spin chains with strong on-site impurities},\ }\href
  {https://doi.org/10.1142/s0129055x22500076} {\bibfield  {journal} {\bibinfo
  {journal} {Reviews in Mathematical Physics}\ }\textbf {\bibinfo {volume}
  {34}},\ \bibinfo {pages} {2250007} (\bibinfo {year} {2022})},\ \Eprint
  {https://arxiv.org/abs/arXiv:2104.00968v1} {arXiv:arXiv:2104.00968v1
  [math-ph]} \BibitemShut {NoStop}%
\bibitem [{\citenamefont {Morningstar}\ \emph {et~al.}(2022)\citenamefont
  {Morningstar}, \citenamefont {Colmenarez}, \citenamefont {Khemani},
  \citenamefont {Luitz},\ and\ \citenamefont {Huse}}]{Morningstar_Huse_2022}%
  \BibitemOpen
  \bibfield  {author} {\bibinfo {author} {\bibfnamefont {A.}~\bibnamefont
  {Morningstar}}, \bibinfo {author} {\bibfnamefont {L.}~\bibnamefont
  {Colmenarez}}, \bibinfo {author} {\bibfnamefont {V.}~\bibnamefont {Khemani}},
  \bibinfo {author} {\bibfnamefont {D.~J.}\ \bibnamefont {Luitz}},\ and\
  \bibinfo {author} {\bibfnamefont {D.~A.}\ \bibnamefont {Huse}},\ }\bibfield
  {title} {\bibinfo {title} {Avalanches and many-body resonances in many-body
  localized systems},\ }\href {https://doi.org/10.1103/PhysRevB.105.174205}
  {\bibfield  {journal} {\bibinfo  {journal} {Phys. Rev. B}\ }\textbf {\bibinfo
  {volume} {105}},\ \bibinfo {pages} {174205} (\bibinfo {year} {2022})},\
  \Eprint {https://arxiv.org/abs/2107.05642} {arXiv:2107.05642
  [cond-mat.dis-nn]} \BibitemShut {NoStop}%
\bibitem [{\citenamefont {Ha}\ \emph {et~al.}(2023)\citenamefont {Ha},
  \citenamefont {Morningstar},\ and\ \citenamefont
  {Huse}}]{Morningstar_Huse_2023}%
  \BibitemOpen
  \bibfield  {author} {\bibinfo {author} {\bibfnamefont {H.}~\bibnamefont
  {Ha}}, \bibinfo {author} {\bibfnamefont {A.}~\bibnamefont {Morningstar}},\
  and\ \bibinfo {author} {\bibfnamefont {D.~A.}\ \bibnamefont {Huse}},\
  }\bibfield  {title} {\bibinfo {title} {Many-body resonances in the avalanche
  instability of many-body localization},\ }\href
  {https://doi.org/10.1103/PhysRevLett.130.250405} {\bibfield  {journal}
  {\bibinfo  {journal} {Phys. Rev. Lett.}\ }\textbf {\bibinfo {volume} {130}},\
  \bibinfo {pages} {250405} (\bibinfo {year} {2023})},\ \Eprint
  {https://arxiv.org/abs/2301.04658} {arXiv:2301.04658 [cond-mat.stat-mech]}
  \BibitemShut {NoStop}%
\bibitem [{\citenamefont {Léonard}\ \emph {et~al.}(2023)\citenamefont
  {Léonard}, \citenamefont {Kim}, \citenamefont {Rispoli}, \citenamefont
  {Lukin}, \citenamefont {Schittko}, \citenamefont {Kwan}, \citenamefont
  {Demler}, \citenamefont {Sels},\ and\ \citenamefont
  {Greiner}}]{Leonard_2023}%
  \BibitemOpen
  \bibfield  {author} {\bibinfo {author} {\bibfnamefont {J.}~\bibnamefont
  {Léonard}}, \bibinfo {author} {\bibfnamefont {S.}~\bibnamefont {Kim}},
  \bibinfo {author} {\bibfnamefont {M.}~\bibnamefont {Rispoli}}, \bibinfo
  {author} {\bibfnamefont {A.}~\bibnamefont {Lukin}}, \bibinfo {author}
  {\bibfnamefont {R.}~\bibnamefont {Schittko}}, \bibinfo {author}
  {\bibfnamefont {J.}~\bibnamefont {Kwan}}, \bibinfo {author} {\bibfnamefont
  {E.}~\bibnamefont {Demler}}, \bibinfo {author} {\bibfnamefont
  {D.}~\bibnamefont {Sels}},\ and\ \bibinfo {author} {\bibfnamefont
  {M.}~\bibnamefont {Greiner}},\ }\bibfield  {title} {\bibinfo {title} {Probing
  the onset of quantum avalanches in a many-body localized system},\ }\href
  {https://doi.org/10.1038/s41567-022-01887-3} {\bibfield  {journal} {\bibinfo
  {journal} {Nature Physics}\ }\textbf {\bibinfo {volume} {19}},\ \bibinfo
  {pages} {481–485} (\bibinfo {year} {2023})},\ \Eprint
  {https://arxiv.org/abs/2012.15270} {arXiv:2012.15270 [cond-mat.quant-gas]}
  \BibitemShut {NoStop}%
\bibitem [{\citenamefont {Szo\l{}dra}\ \emph {et~al.}(2024)\citenamefont
  {Szo\l{}dra}, \citenamefont {Sierant}, \citenamefont {Lewenstein},\ and\
  \citenamefont {Zakrzewski}}]{Sierant_2024_PRB}%
  \BibitemOpen
  \bibfield  {author} {\bibinfo {author} {\bibfnamefont {T.}~\bibnamefont
  {Szo\l{}dra}}, \bibinfo {author} {\bibfnamefont {P.}~\bibnamefont {Sierant}},
  \bibinfo {author} {\bibfnamefont {M.}~\bibnamefont {Lewenstein}},\ and\
  \bibinfo {author} {\bibfnamefont {J.}~\bibnamefont {Zakrzewski}},\ }\bibfield
   {title} {\bibinfo {title} {Catching thermal avalanches in the disordered xxz
  model},\ }\href {https://doi.org/10.1103/PhysRevB.109.134202} {\bibfield
  {journal} {\bibinfo  {journal} {Phys. Rev. B}\ }\textbf {\bibinfo {volume}
  {109}},\ \bibinfo {pages} {134202} (\bibinfo {year} {2024})},\ \Eprint
  {https://arxiv.org/abs/2402.01362} {arXiv:2402.01362 [cond-mat.dis-nn]}
  \BibitemShut {NoStop}%
\bibitem [{\citenamefont {Scocco}\ \emph {et~al.}(2024)\citenamefont {Scocco},
  \citenamefont {Passarelli}, \citenamefont {Collura}, \citenamefont
  {Lucignano},\ and\ \citenamefont {Russomanno}}]{Scocco_2024}%
  \BibitemOpen
  \bibfield  {author} {\bibinfo {author} {\bibfnamefont {A.}~\bibnamefont
  {Scocco}}, \bibinfo {author} {\bibfnamefont {G.}~\bibnamefont {Passarelli}},
  \bibinfo {author} {\bibfnamefont {M.}~\bibnamefont {Collura}}, \bibinfo
  {author} {\bibfnamefont {P.}~\bibnamefont {Lucignano}},\ and\ \bibinfo
  {author} {\bibfnamefont {A.}~\bibnamefont {Russomanno}},\ }\bibfield  {title}
  {\bibinfo {title} {Thermalization propagation front and robustness against
  avalanches in localized systems},\ }\href
  {https://doi.org/10.1103/PhysRevB.110.134204} {\bibfield  {journal} {\bibinfo
   {journal} {Phys. Rev. B}\ }\textbf {\bibinfo {volume} {110}},\ \bibinfo
  {pages} {134204} (\bibinfo {year} {2024})},\ \Eprint
  {https://arxiv.org/abs/2407.20985} {arXiv:2407.20985 [quant-ph]} \BibitemShut
  {NoStop}%
\bibitem [{\citenamefont {Goihl}\ \emph {et~al.}(2019)\citenamefont {Goihl},
  \citenamefont {Eisert},\ and\ \citenamefont {Krumnow}}]{Goihl_2019}%
  \BibitemOpen
  \bibfield  {author} {\bibinfo {author} {\bibfnamefont {M.}~\bibnamefont
  {Goihl}}, \bibinfo {author} {\bibfnamefont {J.}~\bibnamefont {Eisert}},\ and\
  \bibinfo {author} {\bibfnamefont {C.}~\bibnamefont {Krumnow}},\ }\bibfield
  {title} {\bibinfo {title} {Exploration of the stability of many-body
  localized systems in the presence of a small bath},\ }\href
  {https://doi.org/10.1103/PhysRevB.99.195145} {\bibfield  {journal} {\bibinfo
  {journal} {Phys. Rev. B}\ }\textbf {\bibinfo {volume} {99}},\ \bibinfo
  {pages} {195145} (\bibinfo {year} {2019})}\BibitemShut {NoStop}%
\bibitem [{\citenamefont {Potirniche}\ \emph {et~al.}(2019)\citenamefont
  {Potirniche}, \citenamefont {Banerjee},\ and\ \citenamefont
  {Altman}}]{Potirniche_2019}%
  \BibitemOpen
  \bibfield  {author} {\bibinfo {author} {\bibfnamefont {I.-D.}\ \bibnamefont
  {Potirniche}}, \bibinfo {author} {\bibfnamefont {S.}~\bibnamefont
  {Banerjee}},\ and\ \bibinfo {author} {\bibfnamefont {E.}~\bibnamefont
  {Altman}},\ }\bibfield  {title} {\bibinfo {title} {Exploration of the
  stability of many-body localization in $ d > 1 $},\ }\href
  {https://doi.org/10.1103/PhysRevB.99.205149} {\bibfield  {journal} {\bibinfo
  {journal} {Phys. Rev. B}\ }\textbf {\bibinfo {volume} {99}},\ \bibinfo
  {pages} {205149} (\bibinfo {year} {2019})},\ \Eprint
  {https://arxiv.org/abs/1805.01475} {arXiv:1805.01475 [cond-mat.dis-nn]}
  \BibitemShut {NoStop}%
\bibitem [{\citenamefont {Pastur}\ and\ \citenamefont
  {Slavin}(2014)}]{Pastur_2014}%
  \BibitemOpen
  \bibfield  {author} {\bibinfo {author} {\bibfnamefont {L.}~\bibnamefont
  {Pastur}}\ and\ \bibinfo {author} {\bibfnamefont {V.}~\bibnamefont
  {Slavin}},\ }\bibfield  {title} {\bibinfo {title} {Area law scaling for the
  entropy of disordered quasifree fermions},\ }\href
  {https://doi.org/10.1103/PhysRevLett.113.150404} {\bibfield  {journal}
  {\bibinfo  {journal} {Phys. Rev. Lett.}\ }\textbf {\bibinfo {volume} {113}},\
  \bibinfo {pages} {150404} (\bibinfo {year} {2014})},\ \Eprint
  {https://arxiv.org/abs/1408.2570} {arXiv:1408.2570 [quant-ph]} \BibitemShut
  {NoStop}%
\bibitem [{\citenamefont {Abdul-Rahman}\ and\ \citenamefont
  {Stolz}(2015)}]{Stolz_2015}%
  \BibitemOpen
  \bibfield  {author} {\bibinfo {author} {\bibfnamefont {H.}~\bibnamefont
  {Abdul-Rahman}}\ and\ \bibinfo {author} {\bibfnamefont {G.}~\bibnamefont
  {Stolz}},\ }\bibfield  {title} {\bibinfo {title} {{A uniform area law for the
  entanglement of eigenstates in the disordered XY chain}},\ }\href
  {https://doi.org/10.1063/1.4938573} {\bibfield  {journal} {\bibinfo
  {journal} {Journal of Mathematical Physics}\ }\textbf {\bibinfo {volume}
  {56}},\ \bibinfo {pages} {121901} (\bibinfo {year} {2015})},\ \Eprint
  {https://arxiv.org/abs/1505.02117} {arXiv:1505.02117 [math-ph]} \BibitemShut
  {NoStop}%
\bibitem [{\citenamefont {Bardarson}\ \emph {et~al.}(2012)\citenamefont
  {Bardarson}, \citenamefont {Pollmann},\ and\ \citenamefont
  {Moore}}]{Bardarson_2012}%
  \BibitemOpen
  \bibfield  {author} {\bibinfo {author} {\bibfnamefont {J.~H.}\ \bibnamefont
  {Bardarson}}, \bibinfo {author} {\bibfnamefont {F.}~\bibnamefont
  {Pollmann}},\ and\ \bibinfo {author} {\bibfnamefont {J.~E.}\ \bibnamefont
  {Moore}},\ }\bibfield  {title} {\bibinfo {title} {Unbounded growth of
  entanglement in models of many-body localization},\ }\href
  {https://doi.org/10.1103/PhysRevLett.109.017202} {\bibfield  {journal}
  {\bibinfo  {journal} {Phys. Rev. Lett.}\ }\textbf {\bibinfo {volume} {109}},\
  \bibinfo {pages} {017202} (\bibinfo {year} {2012})},\ \Eprint
  {https://arxiv.org/abs/1202.5532} {arXiv:1202.5532 [cond-mat.str-el]}
  \BibitemShut {NoStop}%
\bibitem [{\citenamefont {Serbyn}\ \emph {et~al.}(2013)\citenamefont {Serbyn},
  \citenamefont {Papi\ifmmode~\acute{c}\else \'{c}\fi{}},\ and\ \citenamefont
  {Abanin}}]{Serbyn_2013}%
  \BibitemOpen
  \bibfield  {author} {\bibinfo {author} {\bibfnamefont {M.}~\bibnamefont
  {Serbyn}}, \bibinfo {author} {\bibfnamefont {Z.}~\bibnamefont
  {Papi\ifmmode~\acute{c}\else \'{c}\fi{}}},\ and\ \bibinfo {author}
  {\bibfnamefont {D.~A.}\ \bibnamefont {Abanin}},\ }\bibfield  {title}
  {\bibinfo {title} {Universal slow growth of entanglement in interacting
  strongly disordered systems},\ }\href
  {https://doi.org/10.1103/PhysRevLett.110.260601} {\bibfield  {journal}
  {\bibinfo  {journal} {Phys. Rev. Lett.}\ }\textbf {\bibinfo {volume} {110}},\
  \bibinfo {pages} {260601} (\bibinfo {year} {2013})},\ \Eprint
  {https://arxiv.org/abs/1304.4605} {arXiv:1304.4605 [cond-mat.str-el]}
  \BibitemShut {NoStop}%
\bibitem [{\citenamefont {Gebert}\ and\ \citenamefont
  {Lemm}(2016)}]{Gebert_2016}%
  \BibitemOpen
  \bibfield  {author} {\bibinfo {author} {\bibfnamefont {M.}~\bibnamefont
  {Gebert}}\ and\ \bibinfo {author} {\bibfnamefont {M.}~\bibnamefont {Lemm}},\
  }\bibfield  {title} {\bibinfo {title} {On polynomial {Lieb}–{Robinson}
  bounds for the {XY} chain in a decaying random field},\ }\href
  {https://doi.org/10.1007/s10955-016-1558-0} {\bibfield  {journal} {\bibinfo
  {journal} {Journal of Statistical Physics}\ }\textbf {\bibinfo {volume}
  {164}},\ \bibinfo {pages} {667–679} (\bibinfo {year} {2016})},\ \Eprint
  {https://arxiv.org/abs/1601.03383} {arXiv:1601.03383 [math-ph]} \BibitemShut
  {NoStop}%
\bibitem [{\citenamefont {Nico-Katz}\ \emph {et~al.}(2022)\citenamefont
  {Nico-Katz}, \citenamefont {Bayat},\ and\ \citenamefont
  {Bose}}]{Nico_Katz_2022}%
  \BibitemOpen
  \bibfield  {author} {\bibinfo {author} {\bibfnamefont {A.}~\bibnamefont
  {Nico-Katz}}, \bibinfo {author} {\bibfnamefont {A.}~\bibnamefont {Bayat}},\
  and\ \bibinfo {author} {\bibfnamefont {S.}~\bibnamefont {Bose}},\ }\bibfield
  {title} {\bibinfo {title} {Information-theoretic memory scaling in the
  many-body localization transition},\ }\href
  {https://doi.org/10.1103/PhysRevB.105.205133} {\bibfield  {journal} {\bibinfo
   {journal} {Phys. Rev. B}\ }\textbf {\bibinfo {volume} {105}},\ \bibinfo
  {pages} {205133} (\bibinfo {year} {2022})},\ \Eprint
  {https://arxiv.org/abs/2009.04470} {arXiv:2009.04470 [quant-ph]} \BibitemShut
  {NoStop}%
\bibitem [{\citenamefont {Damanik}\ and\ \citenamefont
  {Tcheremchantsev}(2008)}]{Damanik_2008}%
  \BibitemOpen
  \bibfield  {author} {\bibinfo {author} {\bibfnamefont {D.}~\bibnamefont
  {Damanik}}\ and\ \bibinfo {author} {\bibfnamefont {S.}~\bibnamefont
  {Tcheremchantsev}},\ }\bibfield  {title} {\bibinfo {title} {Quantum dynamics
  via complex analysis methods: General upper bounds without time-averaging and
  tight lower bounds for the strongly coupled fibonacci hamiltonian},\ }\href
  {https://doi.org/https://doi.org/10.1016/j.jfa.2008.08.010} {\bibfield
  {journal} {\bibinfo  {journal} {Journal of Functional Analysis}\ }\textbf
  {\bibinfo {volume} {255}},\ \bibinfo {pages} {2872} (\bibinfo {year}
  {2008})},\ \Eprint {https://arxiv.org/abs/0801.3399} {arXiv:0801.3399
  [math.SP]} \BibitemShut {NoStop}%
\bibitem [{\citenamefont {Kim}\ and\ \citenamefont {Huse}(2013)}]{Kim_2013}%
  \BibitemOpen
  \bibfield  {author} {\bibinfo {author} {\bibfnamefont {H.}~\bibnamefont
  {Kim}}\ and\ \bibinfo {author} {\bibfnamefont {D.~A.}\ \bibnamefont {Huse}},\
  }\bibfield  {title} {\bibinfo {title} {Ballistic spreading of entanglement in
  a diffusive nonintegrable system},\ }\href
  {https://doi.org/10.1103/PhysRevLett.111.127205} {\bibfield  {journal}
  {\bibinfo  {journal} {Phys. Rev. Lett.}\ }\textbf {\bibinfo {volume} {111}},\
  \bibinfo {pages} {127205} (\bibinfo {year} {2013})},\ \Eprint
  {https://arxiv.org/abs/1306.4306} {arXiv:1306.4306 [quant-ph]} \BibitemShut
  {NoStop}%
\bibitem [{\citenamefont {Rakovszky}\ \emph {et~al.}(2019)\citenamefont
  {Rakovszky}, \citenamefont {Pollmann},\ and\ \citenamefont {von
  Keyserlingk}}]{Rakovszky_2019}%
  \BibitemOpen
  \bibfield  {author} {\bibinfo {author} {\bibfnamefont {T.}~\bibnamefont
  {Rakovszky}}, \bibinfo {author} {\bibfnamefont {F.}~\bibnamefont
  {Pollmann}},\ and\ \bibinfo {author} {\bibfnamefont {C.~W.}\ \bibnamefont
  {von Keyserlingk}},\ }\bibfield  {title} {\bibinfo {title} {Sub-ballistic
  growth of {R\'enyi} entropies due to diffusion},\ }\href
  {https://doi.org/10.1103/PhysRevLett.122.250602} {\bibfield  {journal}
  {\bibinfo  {journal} {Phys. Rev. Lett.}\ }\textbf {\bibinfo {volume} {122}},\
  \bibinfo {pages} {250602} (\bibinfo {year} {2019})},\ \Eprint
  {https://arxiv.org/abs/1901.10502v2} {arXiv:1901.10502v2 [cond-mat.str-el]}
  \BibitemShut {NoStop}%
\bibitem [{\citenamefont {$\check{Z}$nidari$\check{c}$}(2023)}]{Znidaric_2023}%
  \BibitemOpen
  \bibfield  {author} {\bibinfo {author} {\bibfnamefont {M.}~\bibnamefont
  {$\check{Z}$nidari$\check{c}$}},\ }\bibfield  {title} {\bibinfo {title}
  {Entanglement growth in diffusive systems},\ }\href
  {https://doi.org/10.1038/s42005-020-0366-7} {\bibfield  {journal} {\bibinfo
  {journal} {Communications Physics}\ }\textbf {\bibinfo {volume} {3}},\
  \bibinfo {pages} {100} (\bibinfo {year} {2023})},\ \Eprint
  {https://arxiv.org/abs/1912.3645} {arXiv:1912.3645 [cond-mat.str-el]}
  \BibitemShut {NoStop}%
\bibitem [{\citenamefont {Rakovszky}\ \emph {et~al.}(2018)\citenamefont
  {Rakovszky}, \citenamefont {Pollmann},\ and\ \citenamefont {von
  Keyserlingk}}]{Rakovszky_2018}%
  \BibitemOpen
  \bibfield  {author} {\bibinfo {author} {\bibfnamefont {T.}~\bibnamefont
  {Rakovszky}}, \bibinfo {author} {\bibfnamefont {F.}~\bibnamefont
  {Pollmann}},\ and\ \bibinfo {author} {\bibfnamefont {C.~W.}\ \bibnamefont
  {von Keyserlingk}},\ }\bibfield  {title} {\bibinfo {title} {Diffusive
  hydrodynamics of out-of-time-ordered correlators with charge conservation},\
  }\href {https://doi.org/10.1103/PhysRevX.8.031058} {\bibfield  {journal}
  {\bibinfo  {journal} {Phys. Rev. X}\ }\textbf {\bibinfo {volume} {8}},\
  \bibinfo {pages} {031058} (\bibinfo {year} {2018})},\ \Eprint
  {https://arxiv.org/abs/1710.09827} {arXiv:1710.09827 [cond-mat.stat-mech]}
  \BibitemShut {NoStop}%
\bibitem [{\citenamefont {Prosen}(2011)}]{Prosen_2011}%
  \BibitemOpen
  \bibfield  {author} {\bibinfo {author} {\bibfnamefont {T.}~\bibnamefont
  {Prosen}},\ }\bibfield  {title} {\bibinfo {title} {Open {$XXZ$} spin chain:
  Nonequilibrium steady state and a strict bound on ballistic transport},\
  }\href {https://doi.org/10.1103/PhysRevLett.106.217206} {\bibfield  {journal}
  {\bibinfo  {journal} {Phys. Rev. Lett.}\ }\textbf {\bibinfo {volume} {106}},\
  \bibinfo {pages} {217206} (\bibinfo {year} {2011})},\ \Eprint
  {https://arxiv.org/abs/1103.1350} {arXiv:1103.1350 [cond-mat.str-el]}
  \BibitemShut {NoStop}%
\bibitem [{\citenamefont {Mahoney}\ and\ \citenamefont
  {Lent}(2024)}]{Mahoney_2024}%
  \BibitemOpen
  \bibfield  {author} {\bibinfo {author} {\bibfnamefont {B.~J.}\ \bibnamefont
  {Mahoney}}\ and\ \bibinfo {author} {\bibfnamefont {C.~S.}\ \bibnamefont
  {Lent}},\ }\bibfield  {title} {\bibinfo {title} {Lieb-robinson correlation
  function for the quantum transverse-field ising model},\ }\href
  {https://doi.org/10.1103/PhysRevResearch.6.023286} {\bibfield  {journal}
  {\bibinfo  {journal} {Phys. Rev. Res.}\ }\textbf {\bibinfo {volume} {6}},\
  \bibinfo {pages} {023286} (\bibinfo {year} {2024})},\ \Eprint
  {https://arxiv.org/abs/2402.11080} {arXiv:2402.11080 [quant-ph]} \BibitemShut
  {NoStop}%
\bibitem [{\citenamefont {Khemani}\ \emph {et~al.}(2015)\citenamefont
  {Khemani}, \citenamefont {Nandkishore},\ and\ \citenamefont
  {Sondhi}}]{Khemani_2015}%
  \BibitemOpen
  \bibfield  {author} {\bibinfo {author} {\bibfnamefont {V.}~\bibnamefont
  {Khemani}}, \bibinfo {author} {\bibfnamefont {R.}~\bibnamefont
  {Nandkishore}},\ and\ \bibinfo {author} {\bibfnamefont {S.~L.}\ \bibnamefont
  {Sondhi}},\ }\bibfield  {title} {\bibinfo {title} {Nonlocal adiabatic
  response of a localized system to local manipulations},\ }\href
  {https://doi.org/10.1038/nphys3344} {\bibfield  {journal} {\bibinfo
  {journal} {Nature Physics}\ }\textbf {\bibinfo {volume} {11}},\ \bibinfo
  {pages} {560–565} (\bibinfo {year} {2015})},\ \Eprint
  {https://arxiv.org/abs/1411.2616} {arXiv:1411.2616 [cond-mat.dis-nn]}
  \BibitemShut {NoStop}%
\bibitem [{\citenamefont {Masanes}(2009)}]{Masanes_2009}%
  \BibitemOpen
  \bibfield  {author} {\bibinfo {author} {\bibfnamefont {L.}~\bibnamefont
  {Masanes}},\ }\bibfield  {title} {\bibinfo {title} {Area law for the entropy
  of low-energy states},\ }\href {https://doi.org/10.1103/PhysRevA.80.052104}
  {\bibfield  {journal} {\bibinfo  {journal} {Phys. Rev. A}\ }\textbf {\bibinfo
  {volume} {80}},\ \bibinfo {pages} {052104} (\bibinfo {year} {2009})},\
  \Eprint {https://arxiv.org/abs/0907.4672} {arXiv:0907.4672 [quant-ph]}
  \BibitemShut {NoStop}%
\bibitem [{\citenamefont {Yin}\ and\ \citenamefont {Lucas}(2023)}]{Yin_2023}%
  \BibitemOpen
  \bibfield  {author} {\bibinfo {author} {\bibfnamefont {C.}~\bibnamefont
  {Yin}}\ and\ \bibinfo {author} {\bibfnamefont {A.}~\bibnamefont {Lucas}},\
  }\bibfield  {title} {\bibinfo {title} {Prethermalization and the local
  robustness of gapped systems},\ }\href
  {https://doi.org/10.1103/PhysRevLett.131.050402} {\bibfield  {journal}
  {\bibinfo  {journal} {Phys. Rev. Lett.}\ }\textbf {\bibinfo {volume} {131}},\
  \bibinfo {pages} {050402} (\bibinfo {year} {2023})},\ \Eprint
  {https://arxiv.org/abs/2209.11242} {arXiv:2209.11242 [cond-mat.str-el]}
  \BibitemShut {NoStop}%
\bibitem [{\citenamefont {Hastings}(2008)}]{Hastings_2008}%
  \BibitemOpen
  \bibfield  {author} {\bibinfo {author} {\bibfnamefont {M.~B.}\ \bibnamefont
  {Hastings}},\ }\bibfield  {title} {\bibinfo {title} {Observations outside the
  light cone: Algorithms for nonequilibrium and thermal states},\ }\href
  {https://doi.org/10.1103/PhysRevB.77.144302} {\bibfield  {journal} {\bibinfo
  {journal} {Phys. Rev. B}\ }\textbf {\bibinfo {volume} {77}},\ \bibinfo
  {pages} {144302} (\bibinfo {year} {2008})},\ \Eprint
  {https://arxiv.org/abs/0801.2161} {arXiv:0801.2161 [quant-ph]} \BibitemShut
  {NoStop}%
\bibitem [{\citenamefont {Foss-Feig}\ \emph {et~al.}(2015)\citenamefont
  {Foss-Feig}, \citenamefont {Gong}, \citenamefont {Clark},\ and\ \citenamefont
  {Gorshkov}}]{Foss-Feig_2015}%
  \BibitemOpen
  \bibfield  {author} {\bibinfo {author} {\bibfnamefont {M.}~\bibnamefont
  {Foss-Feig}}, \bibinfo {author} {\bibfnamefont {Z.-X.}\ \bibnamefont {Gong}},
  \bibinfo {author} {\bibfnamefont {C.~W.}\ \bibnamefont {Clark}},\ and\
  \bibinfo {author} {\bibfnamefont {A.~V.}\ \bibnamefont {Gorshkov}},\
  }\bibfield  {title} {\bibinfo {title} {Nearly linear light cones in
  long-range interacting quantum systems},\ }\href
  {https://doi.org/10.1103/PhysRevLett.114.157201} {\bibfield  {journal}
  {\bibinfo  {journal} {Phys. Rev. Lett.}\ }\textbf {\bibinfo {volume} {114}},\
  \bibinfo {pages} {157201} (\bibinfo {year} {2015})},\ \Eprint
  {https://arxiv.org/abs/1410.3466} {arXiv:1410.3466 [quant-ph]} \BibitemShut
  {NoStop}%
\bibitem [{\citenamefont {Kuwahara}\ and\ \citenamefont
  {Saito}(2020)}]{Kuwahara_2020}%
  \BibitemOpen
  \bibfield  {author} {\bibinfo {author} {\bibfnamefont {T.}~\bibnamefont
  {Kuwahara}}\ and\ \bibinfo {author} {\bibfnamefont {K.}~\bibnamefont
  {Saito}},\ }\bibfield  {title} {\bibinfo {title} {Strictly linear light cones
  in long-range interacting systems of arbitrary dimensions},\ }\href
  {https://doi.org/10.1103/PhysRevX.10.031010} {\bibfield  {journal} {\bibinfo
  {journal} {Phys. Rev. X}\ }\textbf {\bibinfo {volume} {10}},\ \bibinfo
  {pages} {031010} (\bibinfo {year} {2020})},\ \Eprint
  {https://arxiv.org/abs/1910.14477} {arXiv:1910.14477 [quanth-ph]}
  \BibitemShut {NoStop}%
\bibitem [{\citenamefont {Baldwin}\ \emph {et~al.}(2023)\citenamefont
  {Baldwin}, \citenamefont {Ehrenberg}, \citenamefont {Guo},\ and\
  \citenamefont {Gorshkov}}]{Baldwin_2023}%
  \BibitemOpen
  \bibfield  {author} {\bibinfo {author} {\bibfnamefont {C.~L.}\ \bibnamefont
  {Baldwin}}, \bibinfo {author} {\bibfnamefont {A.}~\bibnamefont {Ehrenberg}},
  \bibinfo {author} {\bibfnamefont {A.~Y.}\ \bibnamefont {Guo}},\ and\ \bibinfo
  {author} {\bibfnamefont {A.~V.}\ \bibnamefont {Gorshkov}},\ }\bibfield
  {title} {\bibinfo {title} {Disordered {Lieb}-{Robinson} bounds in one
  dimension},\ }\href {https://doi.org/10.1103/PRXQuantum.4.020349} {\bibfield
  {journal} {\bibinfo  {journal} {PRX Quantum}\ }\textbf {\bibinfo {volume}
  {4}},\ \bibinfo {pages} {020349} (\bibinfo {year} {2023})},\ \Eprint
  {https://arxiv.org/abs/2208.05509} {arXiv:2208.05509 [cond-mat.dis-nn]}
  \BibitemShut {NoStop}%
\bibitem [{\citenamefont {Tran}\ \emph {et~al.}(2019)\citenamefont {Tran},
  \citenamefont {Guo}, \citenamefont {Su}, \citenamefont {Garrison},
  \citenamefont {Eldredge}, \citenamefont {Foss-Feig}, \citenamefont {Childs},\
  and\ \citenamefont {Gorshkov}}]{Tran_2019}%
  \BibitemOpen
  \bibfield  {author} {\bibinfo {author} {\bibfnamefont {M.~C.}\ \bibnamefont
  {Tran}}, \bibinfo {author} {\bibfnamefont {A.~Y.}\ \bibnamefont {Guo}},
  \bibinfo {author} {\bibfnamefont {Y.}~\bibnamefont {Su}}, \bibinfo {author}
  {\bibfnamefont {J.~R.}\ \bibnamefont {Garrison}}, \bibinfo {author}
  {\bibfnamefont {Z.}~\bibnamefont {Eldredge}}, \bibinfo {author}
  {\bibfnamefont {M.}~\bibnamefont {Foss-Feig}}, \bibinfo {author}
  {\bibfnamefont {A.~M.}\ \bibnamefont {Childs}},\ and\ \bibinfo {author}
  {\bibfnamefont {A.~V.}\ \bibnamefont {Gorshkov}},\ }\bibfield  {title}
  {\bibinfo {title} {Locality and digital quantum simulation of power-law
  interactions},\ }\href {https://doi.org/10.1103/PhysRevX.9.031006} {\bibfield
   {journal} {\bibinfo  {journal} {Phys. Rev. X}\ }\textbf {\bibinfo {volume}
  {9}},\ \bibinfo {pages} {031006} (\bibinfo {year} {2019})},\ \Eprint
  {https://arxiv.org/abs/1808.05225} {arXiv:1808.05225 [quant-ph]} \BibitemShut
  {NoStop}%
\end{thebibliography}%

\end{document}